\documentclass[%
 twocolumn, 12pt,
 amssymb,
 aps,
]{aastex631}

\usepackage{graphicx}
\usepackage{dcolumn}
\usepackage{bm}
\usepackage{floatrow}
\usepackage{hyperref}
\usepackage{xcolor}

\usepackage{graphicx}
\usepackage{natbib}
\usepackage{subcaption}
\usepackage[fleqn]{amsmath}
\begin{document}

\title{Virgo Filaments VIII: Characterizing the Structural Parameters of Virgo Galaxies with Machine Learning to Probe Environmental Quenching}

\author[0009-0005-0303-0330]{Kim Conger}
\altaffiliation[Email: ]{kconger@ku.edu}
\affiliation{University of Kansas, Department of Physics and Astronomy, 1251 Wescoe Hall Drive, Room 1082, Lawrence, KS 66049, USA\\}

\author[0000-0001-5851-1856]{Gregory Rudnick}
\affiliation{University of Kansas, Department of Physics and Astronomy, 1251 Wescoe Hall Drive, Room 1082, Lawrence, KS 66049, USA\\}

\author[0000-0001-8518-4862]{Rose A. Finn}
\affiliation{Department of Physics and Astronomy, Siena University, 515 Loudon Road, Loudonville, NY 12211, USA\\}

\author[0000-0002-3144-2501]{Rebecca A. Koopmann}
\affiliation{Department of Physics \& Astronomy, Union College, Schenectady, NY, 12308, USA\\}

\author[0000-0003-0980-1499]{Benedetta Vulcani}
\affiliation{INAF- Osservatorio astronomico di Padova, Vicolo Osservatorio 5, I-35122 Padova, Italy\\}

\author[0000-0002-5177-727X]{Dennis Zaritsky}
\affiliation{Steward Observatory, University of Arizona, 933 North Cherry Avenue, Tucson, AZ 85721-0065, USA\\}

\author[0000-0001-6831-0687]{Gianluca Castignani}
\affiliation{INAF - Osservatorio di Astrofisica e Scienza dello Spazio di Bologna, via Gobetti 93/3, I-40129, Bologna, Italy\\}

\author[0000-0003-2658-7893]{Francoise Combes}
\affiliation{Observatoire de Paris, LUX, Collège de France, CNRS, PSL University, Sorbonne University, 75014, Paris\\}

\author[0000-0002-6220-9104]{Gabriella De Lucia}
\affiliation{INAF -- Osservatorio Astronomico di Trieste, Via Tiepolo 11, I-34131 Trieste, Italy\\}

\author{Matteo Fossati}
\affiliation{Dipartimento di Fisica G. Occhialini, Universit\`a degli Studi di Milano-Bicocca, Piazza della Scienza 3, I-20126 Milano, Italy}

\author{Gautam Nagaraj}
\affiliation{Laboratoire d'Astrophysique, \'Ecole Polytechnique F\'ed\'erale de Lausanne (EPFL), Route de la Sorge, 1015 Lausanne, Switzerland}

\author[0009-0001-1809-4821]{Daria Zakharova}
\affiliation{INAF -- Osservatorio Astronomico di Trieste, Via Tiepolo 11, I-34131 Trieste, Italy\\}

\date{\today}

\begin{abstract}
\noindent 
A persistent challenge in galaxy evolution involves disentangling the many correlated properties in order to isolate the effects of a galaxy's environment on its star formation history. To address this multidimensionality problem, we apply k-means clustering to 2831 galaxies in the extended regions around the Virgo cluster to define objective, reproducible subsets of structurally similar galaxies. Using measurements of size, light distribution, and stellar mass, k-means partitions these galaxies into three feature classes (FCs): dwarfs, spheroids, and large disks. In addition to being structurally different, these FCs show distinct offsets from the star-forming main sequence to $>3\sigma$ significance, with the spheroid population systematically shifted to lower star formation rates. Examining environmental dependence within each FC, we find that denser environments are associated with progressively stronger quenching. However, star formation for the dwarf and large disk galaxies is not strongly affected until the rich group and cluster environments. For the spheroid galaxies, star formation instead smoothly decreases as environment density increases. We verify that these trends are not driven by differences in Sersic index within each environment, suggesting that the effectiveness of environmental quenching depends on the galaxy's structural class. Our results speak to the utility of a simple machine learning model to create broad classes of structurally similar galaxies based on a small set of parameters, which has important implications for navigating this data rich era of astronomy.
\end{abstract}


\section{Introduction}
\label{sec:introduction}

One of the unifying pillars in astronomy is the idea that, on scales less than tens to hundreds of megaparsecs (Mpc), matter in the Universe is distributed heterogeneously \citep{tifft1976, joeveer1978, sousbie2011A, tully2014, kuchner2022, wang2023, hatamnia2026}. Known colloquially as the cosmic web \citep{bond1996}, the architecture of this large scale structure comprises dense nodes of groups and clusters, filaments that string together and feed these nodes, and sparse regions, sometimes referred to as ``the field."

Many centuries of galaxy evolution studies have also revealed that galaxies are at the mercy of their environment, which corresponds to their current place in the cosmic web. Galaxies in cluster and group environments tend toward lower star formation rates (SFR) \citep[e.g.][]{balogh1997, vulcani2010} and gas contents \citep[e.g.][]{chung2009, scott2013, boselli2014}, smaller star-forming disks \citep[e.g.][]{koopmann2004, schaefer2017, finn2018, conger2025}, and earlier type morphologies \citep[e.g.][]{dressler1980, postman1984, goto2003,vulcani2023}. The physical mechanisms processing these galaxies range from the hydrodynamic removal of cold gas from the disks of infalling galaxies \citep[ram pressure stripping;][]{gunn1972, quilis2000, fosseti2018, tonnesen2019} to the decoupling of a galaxy from its hot gas halo before the gas can sufficiently cool for star formation \citep[``strangulation;"][]{larson1980}, to galaxy mergers \citep{toomre1972}. 

Meanwhile, many of these same consequences of environmental processing also correlate more broadly with traditional galaxy morphologies. Galaxies generally fall into two categories, spheroid-dominated early-type galaxies (ETGs) and disk-dominated late-type galaxies (LTGs), with irregular and transitional shapes scattered throughout \citep[e.g.][]{hubble1926, devaucouleurs1959}. These categories correlate with different physical properties. ETGs, compared to their LTG counterparts, have lower SFR \citep[e.g.][]{balogh1997, vulcani2010, jeong2022, ribeiro2023, williams2025}, lower dust and gas contents \citep[e.g.][]{smith2012, paspaliaris2023, ruffa2024, xiao2025}, and are redder in color due to their older stellar populations \citep[e.g.][]{visvanathan1977, renzini2006} \textendash{} all characteristics hinted at by their ``red and dead" designation among astronomers. 

Because the morphology used to define galaxy classes correlates with star formation activity, establishing whether trends with SFR between these classes are tied to morphology, environment or both becomes complicated. Stated differently, galaxies exist in a multidimensional feature space in which many of these features are covariant, such as morphology and SFR \citep[e.g.][]{kennicutt1998, conselice2014, cano2019}. Research aiming to isolate the role of environment on star formation must choose which galaxy features to control for and how, and which to measure. One possible choice to simplify the problem is to use the classic bimodal description of ETGs and LTGs. However, this classification scheme is not independent of star formation; many visual features used to assign galaxy labels, such as clumpiness, asymmetry, and well-defined spiral arms, are themselves enhanced by young stellar populations. 


Rather than rely on morphology based on traditional classifications, a second strategy is to use parametric structural parameters, such as stellar disk size and the concentration of stellar light. Since ram pressure and strangulation act on a galaxy's gas content and star formation activity while generally only having secondary effects on its pre-existing stellar mass distribution \citep[e.g.][]{moss2000, koopmann2006, gavazzi2010, finn2018, tonnesen2019, conger2025, vulcani2026}, a galaxy's stellar structure is well-suited for creating controlled galaxy populations and isolating environmental processing effects. That is, with structurally similar galaxy classes we can more directly link environmental effects and a galaxy's star formation efficiency.


Opting for structural properties, however, does not erase the problem of having to pick where to divide galaxies in feature space to define distinct subsets. These subjective decisions only become more numerous with more structural features, which necessarily reduces the reproducibility and consistency between results. Modern datasets like SDSS \citep{sdss_dr10_2014}, DESI \citep{desi2025}, The Vera Rubin Observatory, and soon the Nancy Grace Roman Space Telescope \citep{roman2024}, amplify this challenge by delivering terabytes of galaxy data per survey. The task of controlling for these innumerable measurements is therefore a difficult venture both within and across research projects, which has made the search for robust, automated techniques all the more pressing.


With the growing prominence of machine learning as a data analysis tool in astrophysics, many studies have tested its potential for organizing galaxy populations into physically meaningful families while respecting the high dimensionality of the data. Supervised machine learning trains models on datasets with predefined labels and learns its own complex decision boundaries across many parameters in a way that is both reproducible and scalable to larger galaxy surveys \citep[e.g.][]{lin2021, khramtsov2022, cao2024, mazoochi2026}. However, this umbrella of techniques is still often tethered to human-derived parameter partitions and morphology definitions. A second consideration is that supervised machine learning model robustness depends in part on how representative the training set is to the sample to which the model will be applied. For example, if the model is only trained on nearby galaxies, the model may not apply as reliably for galaxies at a higher redshift.

In contrast, unsupervised machine learning (UML) algorithms can determine class labels based on the underlying distribution of the data rather than on a priori morphological definitions or hard cutoff thresholds. Many studies have incorporated versions of UML to investigate data-driven morphology classifications. K-means clustering, for instance, is a common approach to partitioning data into similar groupings according to the data's distribution in multidimensional space. One appeal of this algorithm is its simplicity -- rather than there being an abundance of hyperparameters, the only knob to tune is $k$, which represents the number of groupings into which k-means will divide the data. Multiple studies have embraced clustering algorithms as their central method for organizing their galaxies into morphology classes \citep[e.g.][]{martin2020, cheng2021, fang2026}. In each of the cited cases, however, UML clustering is a step in a much more complicated pipeline. For instance, all three projects involve their own ML-based strategies for extracting many galaxy features, which may or may not be directly connected to the galaxy's SFR. Their choices of $k$, ranging from 20 to 160, are also intended to tease out subtle groupings of similar galaxies in both structure and star formation; however, two of the three papers end up still using visual inspection to place these finely-split galaxy ``clusters" into broad morphology classes \citep{martin2020, fang2026}.

While these multi-step classification pipelines are suitable for the authors' specialized science cases, this paper will explore whether a more simplified application of k-means clustering has a place in the comprehensive ladder of UML techniques. The appeal of this simplicity is not only to facilitate the interpretation of the results, but also to provide a reproducible framework for defining controlled galaxy classes across large datasets where manual inspection is intractable. Within the context of the Virgo Filaments project \citep{castignanivirgo1}, we then assess whether this frame can not only recover meaningful environment trends without the complexity of more specialized UML pipelines, but also uncover new trends otherwise hidden. 

In this work, we directly apply k-means to a set of 2831 galaxies in and surrounding the Virgo cluster (0.002$ < $z$ < $0.011). The clustering acts on a low-dimensional space composed of only four structural properties, which originate directly from parametric model fitting of either the galaxy 2D profile or its spectral energy distribution (SED). Since we do not expect global environment to quickly alter the underlying stellar structure of a galaxy \citep[e.g.][]{moss2000, koopmann2006, finn2018, tonnesen2019, conger2025}, our clustering algorithm partitions the data into broadly structurally similar galaxy populations. This organization allows us to robustly isolate environmental effects on star formation and related galaxy properties across these populations.

Section \ref{sec:data} will introduce our image sources, our environment labels as defined in \citet{castignani2022}, and the modeling tools we use to generate single S\'ersic fits and SED profiles of our galaxies. In Section \ref{sec:sample}, we describe the various quality checks and completeness limits applied to the Virgo catalog to create a UML-ready subset. We provide an overview of our k-means pipeline in Section \ref{sec:clustering} along with the resultant Feature Classes (FCs), and we ascribe physical meaning to these FCs in Section \ref{sec:results}. Our paper discusses interpretations of our FC feature distributions in light of existing observational and theoretical contexts in Section \ref{sec:discussion}. We conclude with a summary of our findings in Section \ref{sec:conclusion}, along with future prospects that concern the application of this work to large scale galaxy surveys.




Throughout the paper, we assume a standard flat $\Lambda$CDM cosmology with $\Omega_M=0.286$ \citep[WMAP 9-year results][]{wmap2013} and a Hubble constant of $H_0$ = $100 h$ km s$^{-1}$ Mpc$^{-1}$, where $h = 0.74$ \citep[e.g.][]{tully2008, riess2019}. We also report AB magnitudes and assume a Chabrier initial mass function (IMF) template \citep{chabrier2003}.

\section{Data}
\label{sec:data}

\subsection{Optical \& \textit{IR} Imaging}

We use imaging data from the DESI Legacy Imaging Survey Data Release 9 \citep[][]{dey2019} and the Wide-Field Infrared Survey Explorer \citep[WISE;][]{wright2010}. The Legacy Survey is a compilation of data products from three telescopes which imaged 14,000 deg$^2$ of the northern hemisphere sky: the Blanco telescope at the Cerro Tololo Inter-American Observatory (The Dark Energy Camera Legacy Survey (DECaLS)); the Mayall Telescope at the Kitt Peak National Observatory (in particular: the Mayall z-band Legacy Survey - MzLS); and the University of Arizona Steward Observatory 2.3 m Bart Bok Telescope, also at Kitt Peak National Observatory (Beijing-Arizona Sky Survey (BASS)). The infrared images are taken from unWISE specifically, which incorporates corrections for spatially-resolved PSFs and background oversubtractions \citep{lang2014, unwise_code_2019}.

\subsection{Virgo Filament Survey}

Our parent sample comprises 6,780 galaxies within and surrounding the Virgo cluster as cataloged by The Virgo Filament Survey \citep[VFS;][]{castignani2022}. The VFS extends 12 virial radii (24 Mpc) in projection from the Virgo cluster center ($\alpha=187^\circ.70$, $\delta=12^\circ.34$, J200) and captures galaxies within a heliocentric velocity range of $500<v_r<3300$ km/s. The lower bound limits contamination from very nearby galaxies while the upper bound is meant to fully include filament systems that may be situated behind Virgo \citep[e.g.,][]{mei2007}.

\subsubsection{Environment Characterization}
\label{sec:env}

We define the global environments of our galaxies according to the following labels: cluster, rich group, poor group, filament, and pure field. These environment assignments are made after correcting the full sample for cosmic flow velocity \citep{mould2000} and redshift-independent distance measurements from the NASA/IPAC Extragalactic Database \citep[NED;][]{steer2017}. A comprehensive review of these corrections and classifications is detailed in \citet{castignani2022} and summarized here. 

For group galaxies, membership is taken from the \citet{kourkchi2017} environment catalog, which identifies  galaxy groups within $v_r<3500$ km s$^{-1}$. We adopt the Kourkchi \& Tully division of rich and poor groups, where rich groups are defined as consisting of five or more members and poor groups contain between two and five members \citep{kourkchi2017}. Filaments were identified using two complementary approaches. First, the VFS was matched to existing catalogs of the Northern Hemisphere from \citet{tully1982} and \citet{kim2016}, yielding eight previously known filaments. An additional new five filaments were identified through visual inspection of high-density contrast spines in 3D supergalactic space. Galaxies within 2 Mpc of a spine were considered members of that filament.

The cluster environment is defined as a spherical region in the 3D supergalactic frame with radius 3.6$h^{-1}$ Mpc ($h=0.74$) centered on Virgo. Galaxies within this radius are classified as cluster members. Finally, pure field galaxies are those which do not belong to any of the other four environments.

Environment membership in the VFS is not mutually exclusive. Galaxies belonging to a filament may simultaneously belong to a rich group, poor group, or the cluster. We accommodate this in our treatment of environments, except for cases where there are too few galaxies to divide into more detailed categories. We order our labels according to density as in \citet{castignani2022}, though we note that relative ordering of poor groups and filaments reflects a practical decision than a strict density ranking.

\subsection{\texttt{GALFIT} Parameters}

We follow the methodology from \citet{conger2025}, which we summarize here.
We use \texttt{GALFIT} \citep{peng2010} to model our galaxies as a single S\'ersic component.  These profiles have seven free parameters: the central x and y pixel (xc, yc), magnitude (m), effective radius (R$_e$), S\'ersic index (\textit{n}), axis ratio (BA), and position angle (PA). The algorithm will iterate over initial guesses of these parameters until finding a local $\chi^2$ minimum in parameter space, constituting a best-fit solution for the galaxy image. As part of \cite{conger2025}, we visually inspected each mask product to verify the absence of shredding of well-resolved massive galaxies and the correct masking of prominent foreground objects. The point spread functions (PSFs) which we use for convolution originate from either the unWISE catalog \citep{schlafly2019} or from the Siena Galaxy Atlas (SGA-2020) custom photometry \citep{moustakas2023}.

\subsection{Star Formation Rate \& Stellar Mass}
\label{sec:CIGALE}

To extract internally consistent stellar mass and SFR estimates for our galaxies, we first construct their observed SEDs using flux measurements at an array of wavelengths, from FUV \citep[GALEX]{GildePaz2007} to optical \citep[Legacy Survey]{dey2019} to 22$\micron$ \citep[WISE]{wright2010}. The procedure for extracting these fluxes involves custom elliptical aperture photometry, detailed in \citet{moustakas2023}, measured within an aperture whose semi-major axis is 1.5 times the galaxy's estimated size\footnote{These sizes are determined using the second moment of the galaxy's light distribution, which is in turn measured after masking out any extraneous sources of flux (e.g., stars) in the image.}. Our fluxes then undergo galactic extinction corrections using the reddening map from \citet{Schlegel1998} and convert to the corresponding extinction for each filter. We transform the extinction corrections to each filter, either using the procedure from the Legacy Survey \citep{dey2019} for the optical \textit{grz} and WISE filters, or \citet{Wyder2007} for the GALEX FUV and NUV filters. 

The Code Investigating GALaxy Emission (\texttt{CIGALE}) package is a python-based SED fitting routine which generates model SED fits to observed fluxes from far-UV to radio, outputting a set of estimated physical properties of galaxies \citep{cigale2019}. We detail our choice of setup modules in Appendix \ref{appendix:cigale}.

Of the galaxies involved in our final k-means sample, 19\% have \texttt{CIGALE} log(SFR) $<$ -3, which is the threshold below which we do not trust the quantitative SFR values. We find that setting a log(SFR)$=-3$ floor to galaxies below this limit negligibly effects the validity or statistical significance of any analysis involving stellar mass and SFR. Nevertheless, we explicitly mark such galaxies in any figures directly plotting an SFR quantity. We revisit this discussion in Section \ref{sec:logsfr}.

\section{Sample Selection}
\label{sec:sample}

Not all of the members of the Virgo parent sample are suitable to present to an UML algorithm. This section addresses the various cuts and conditions we impose on the parent sample.

\subsection{Photometry Quality Flags}

We adopt the W1 and g-band model parameters only as these wavelengths provide complementary measures of stellar structure: whereas W1 traces the underlying stellar mass of a galaxy, \textit{g}-band is more sensitive to recent star formation. The global signal-to-noise ratio (S/N) of a galaxy dictates to first order the reliability of its measured structural parameters, so we require galaxies to have an S/N $\ge$ 20 in the W1 band \citep[e.g.][]{vanderwel2012}. There are no galaxies in our sample that have \textit{g}-band SNR$\le$ 20 that also have W1 S/N $\ge$ 20, so applying the same S/N $\ge$ 20 for \textit{g}-band is redundant. We also remove galaxies with bright foreground stars \citep{moustakas2023}. The S/N cut alone removes 481 (7\%) of the total VFS sample, with 114 (2\%) more galaxies removed using the star flag.

Closely tied to photometry quality, the \texttt{GALFIT} quality criterion encompasses the overall robustness of the galaxy's best-fit S\'ersic profile in both \textit{g}-band and W1. \texttt{GALFIT} outputs a numerical error in cases where one or more model parameters are not reliable, which allows us to easily filter out these failure modes. We also restrict model S\'ersic indices to $n \leq 6$, as higher values indicate a flux profile contaminated by a poorly-masked nearby galaxy or star, or a bright central point source. Combined, these \texttt{GALFIT} cuts remove an additional 971 (14\%) galaxies, of which 778 are W1 failure modes.

The next cut is to axis ratios (B/A), defined as the ratio between the semi-minor and semi-major axes of the model galaxy profile. In the Local Cluster Survey, \cite{finn2023} found that there is a weak yet existing correlation between galaxy ellipticity and \texttt{GALFIT} effective radius. The authors control for this link by requiring galaxies have an $> 0.25$ B/A. Elliptical galaxies with B/A $<0.25$ morphologies are rare \citep[e.g.,][]{padilla2008, rodriguez2013} and we assume that disk galaxies in our sample are intrinsically circular, so this ratio roughly corresponds to an inclination of 75.5$^\circ$. We remove 375 (5\%) galaxies after imposing this cut.


\subsubsection{Mass Completeness Limit}

To ensure the range of stellar masses is equally complete across the full range of redshifts in our galaxy sample, we adapt a mass completeness technique described in \citet{marchesini2009} and \citet{rudnick2017}. The aim of this procedure is to estimate the minimum stellar mass a galaxy at the far end of our catalog's redshift range must have in order to be detected regardless of its observed mass-to-light ratio (M$_*$/L$_{r}$). Using the full VFS, we first isolate the galaxies around this maximum redshift, z$_{max}$ $\sim$ 0.015. From this group, we choose the galaxies brighter than our m$_r$ limit by 0.5-1 mag. This magnitude limit is taken from the Sloan Digital Sky Survey spectroscopic limit of m$_r$ = 17.77 \citep{sdss_dr10_2014}. Our bright subsample at the end of our redshift distribution is far enough above the magnitude completeness limit to be equally complete for the observed range of M$_*$/L$_{r}$. We then fade the $r$-band luminosities of these galaxies and correspondingly reduce their stellar masses such that they now lie at the m$_r$ limit while preserving their M$_*$/L$_{r}$. That is, (M$_{scaled}$/L$_{scaled}$) = (M$_*$/L$_{r}$), with L$_{scaled}$ chosen such that the galaxy's luminosity is now at the m$_r$ limit and M$_{scaled}$ is altered to ensure that the ratios remain unchanged.



In other words, the result is a synthetic collection of galaxies scaled to the m$_r$ limit and still covering the full range of M$_*$/L$_{r}$. We then locate the upper 95th percentile of M$_{scaled}$, which now corresponds to approximately the highest M$_*$/L$_{r}$ due to a roughly constant L$_{scaled}$ at the m$_{r}$ limit. This M$_{scaled}$ therefore gives the highest observed M$_*$/L$_{r}$ ratio at which we can still detect VFS galaxies at the highest sample redshift and at our apparent magnitude limit. We find this completeness limit to be $\log{M_*} = 8.15$, which excludes 2040 (30\%) more galaxies.

Following this full spectrum of quality checks, our final sample comprises 2870 galaxies in and within the vicinity of the Virgo cluster. 

\section{K-Means Clustering}
\label{sec:clustering}
To partition our galaxies in N-dimensional space into structurally similar classes, we use k-means clustering. As introduced in Section \ref{sec:introduction}, k-means is a UML method that will create classes of similar data points in N-dimensional space depending on the desired number of classes (k). The algorithm iterates over random guesses for the k centroid coordinates until the average intra-cluster distance between the data and their cluster centroid reaches a minimum. We explain in this section our choices for the two clustering dials in our control: the input features N and the desired k.


\subsection{Feature Selection}

We choose four galaxy parameters for our clustering algorithm: stellar mass, \textit{g}-band R$_e$, and \textit{g}-band and W1 \textit{n}. These four parameters in tandem provide easily interpretable descriptions of galaxy structure and light distribution without overcomplicating the dimensionality of our feature space. While stellar mass captures the overall stellar content in a galaxy, \textit{g}-band and W1 provide complementary measures of the younger and older populations of stars, respectively. These band measurements come from the available \texttt{GALFIT} models for the VFS \citep{conger2025}, for which we convert R$_e$ from pixels to kiloparsecs (kpc) using the cosmic flow velocities from \citet{castignani2022}. 

\begin{figure}[h]
\includegraphics[scale=0.54]{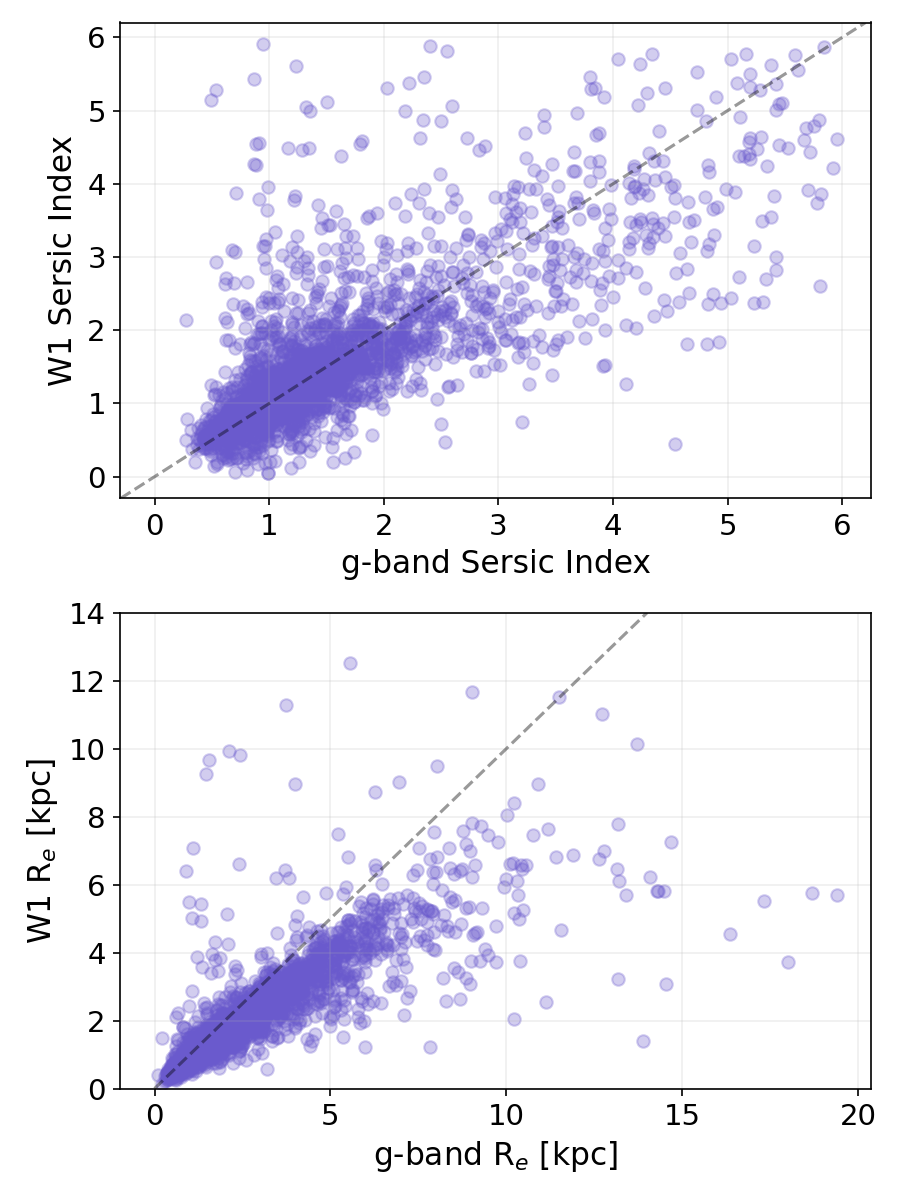}
\centering
\caption{A comparison of the \textit{n} (top) and R$_e$ (bottom) values for the \textit{g}-band and W1 \texttt{GALFIT} models. The 1-to-1 dashed lines are shown for comparison. The \textit{n} values show a high degree of scatter, motivating the use of both the \textit{g}-band and W1 \textit{n} in the subsequent analysis.}
\vspace{.5mm}
\label{fig:corr}
\end{figure}

While we include both R$_e$ and \textit{n} from the \textit{g}-band models, we only include the \textit{n} from W1. This decision is due to a strong correlation between the W1 and \textit{g}-band effective radii. In Figure \ref{fig:corr}, we plot the R$_e$ and \textit{n} for both bands, with a dashed 1-to-1 line for reference. Whereas both panels show strong correlations, the \textit{n} data have much more scatter ($\rho$ = 0.78, where $\rho$ is the Spearman rank coefficient) whereas the R$_e$ are more tightly correlated ($\rho$ = 0.92) with a systematic deviation toward lower W1 R$_e$ for higher \textit{g}-band R$_e$. 


An important consideration with k-means clustering is the scaling of the input parameters. Since k-means uses Euclidean distances to assign points to centroids, features with larger values or numerical ranges (e.g., R$_e$) are weighted more heavily than those with with smaller values or ranges (e.g., \textit{n}). We standardize all input features using the \texttt{StandardScaler} function from \texttt{sklearn.preprocessing}, which rescales each feature to have zero mean and unit variance. This transformation is written as
\begin{gather*}
    x' = \frac{x-\mu}{\sigma}
\end{gather*}
and ensures that no single parameter dominates the clustering solely due to differences in scale.


\subsubsection{Active Galactic Nuclei}
\label{sec:AGN}

We assess whether our results vary with the inclusion or exclusion of galaxies hosting active galactic nuclei (AGN). As in \citet{conger2025}, we flag AGN either through their WISE colors \citep[IR-luminous AGN; e.g.][]{assef2010, asmus2020} or optical emission line ratios \citep[BPT AGN; e.g.][]{kauffmann2003, kewley2004}. We find that $<$1\% of each galaxy class were IR-luminous AGN. In all instances, the class distributions of AGN galaxies plotted on SFR-M$_\ast$ axes follow the general class populations. Moreover, the removal of AGN either before or after applying k-means clustering only affects our results within the uncertainties cited when AGN galaxies are included. We elect to keep these galaxies in our full sample for the purposes of strengthening our number statistics.


\section{Results}
\label{sec:results}


\subsection{Feature Class Creation}

With our sample of 2870 galaxies, we run the k-means algorithm on our scaled \texttt{GALFIT} parameters, choosing k=3 cluster centroids. We set a random state for reproducibility, though we find that varying the random state does not affect the results beyond the uncertainties shown in this section. Specifically, we run our algorithm 100 times with different random states for each run and 99\% of our sample is consistently assigned to the same FC. This consistency is largely due to the stability of the centroids at k=3 being higher than for any alternative k value tested, with the exception of k=2 (see Appendix \ref{appendix:stability}).

\begin{figure*}[t]
\centering
\includegraphics[width=\linewidth]{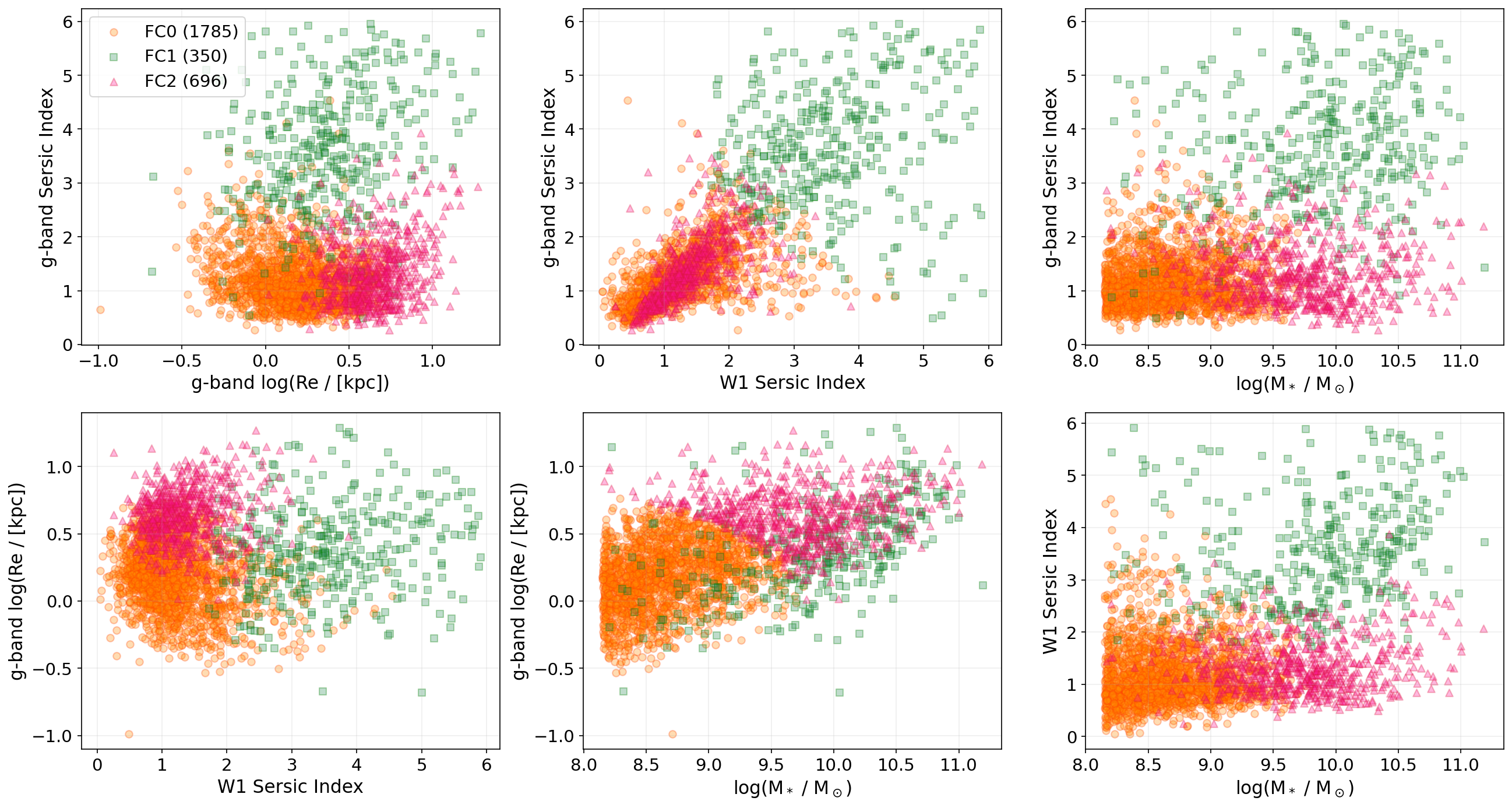}
\caption{2D projections of our k-means classes on physical axes. The axes are every combination of our four input features: W1 and \textit{g}-band \textit{n}, \textit{g}-band R$_e$, and stellar mass. The marker designs correspond to our three classes: FC0 (orange circles), FC1 (green squares), and FC2 (magenta triangles). The number of galaxies comprising each group is given parenthetically in the legend.} 

\label{fig:faux_pca}
\end{figure*}

To visualize the FC partitions, we plot our clustered galaxies in Figure \ref{fig:faux_pca} on six sets of physical axes, one for each combination of our four input features. In each of these projections, the galaxy sample forms three broad classes that are either broadly distinguishable or intertwined. While each FC occupies a different region in nearly all but the middle column plots, there is significant overlap of the data in all cases. 

Nevertheless, within these FCs we observe patterns emerging, illustrating the utility of applying k-means to a multidimensional space. The three distinct FCs are labeled in the legend of Figure \ref{fig:faux_pca} along with the number of members. Starting with FC0, the galaxies generally lie toward lower \textit{n}, \textit{g}-band R$_e$, and stellar mass. FC2 has comparable \textit{n} to FC0 but with larger R$_e$ and stellar mass. The galaxies of FC1, meanwhile, tend to have the highest \textit{n} in both bands, low \textit{g}-band R$_e$, and span the full range of stellar masses in our sample. We expand further on the structural parameter distributions within each FC in Section \ref{sec:strucparams}.



\subsubsection{Outlier Detection \& Visual Inspection}
\label{sec:IQR}

After generating our subset using the selections described above, we evaluated the output of a preliminary run of our clustering algorithm. We noticed that there were galaxies far beyond the cluster centroids, and in some cases the outliers were even being marked as cluster centroids. We identified these outliers using interquartile range (IQR) clipping, which expands the 25\% and 75\% percentiles by some scale factor and labels the points beyond this bloated distribution as outliers. We choose a scale factor of 2 and create a pool of 41 outlier galaxies that required additional screening. In particular, we visually inspected the optical image, W1 image, \textit{g}-band and W1 \texttt{GALFIT} models and residuals, and the best-fit parameters.

After inspection, we found that 35 of these 41 galaxies were in fact attributable to unreliable \texttt{GALFIT} models that managed to bypass our quality checks. Problems ranged from bright foreground stars not included in the SGA-2020 bitmask flags, to nontrivial offsets of the model center from the galaxy. There was even one instance of the model's W1 R$_e$ being larger than the width of the postage stamp, a width which we chose based on a multiple of the central galaxy's isophotal radius. We remove these 32 objects from our sample. Most of the these vetted galaxies had W1 $n<0.1$ or small W1 R$_e$, which we later confirmed to be diffuse dwarf galaxies. We apply a secondary diagnostic to the remaining galaxies, using the size-stellar mass distribution for low redshift objects from \citet{vanderwel2014}. We found two galaxies with unphysical \textit{g}-band R$_e$ for their stellar mass. In both cases, these galaxies were well resolved and had a \textit{g}-band \texttt{GALFIT} profile that traced a prominent central feature not aligned with the semi-major axis of the disk, resulting in an optical size of $>$30 kpc. We also remove these two galaxies from the sample.

Even if this set of 41 outliers comprises a small percentage of the total Virgo galaxies, we note that k-means clustering is a supplementary quality check of the data to detect objects that may require visual followup. Fascinatingly, among the outliers were galaxies whose \texttt{GALFIT} models were spurious due to a starkly asymmetric profile, suggesting that this technique of outlier detection also captures objects of scientific interest. 

\subsection{Feature Properties of Each Feature Class}
\label{sec:strucparams}

\begin{figure*}
\begin{subfigure}{0.45\textwidth}
\centering
\includegraphics[width=\linewidth]{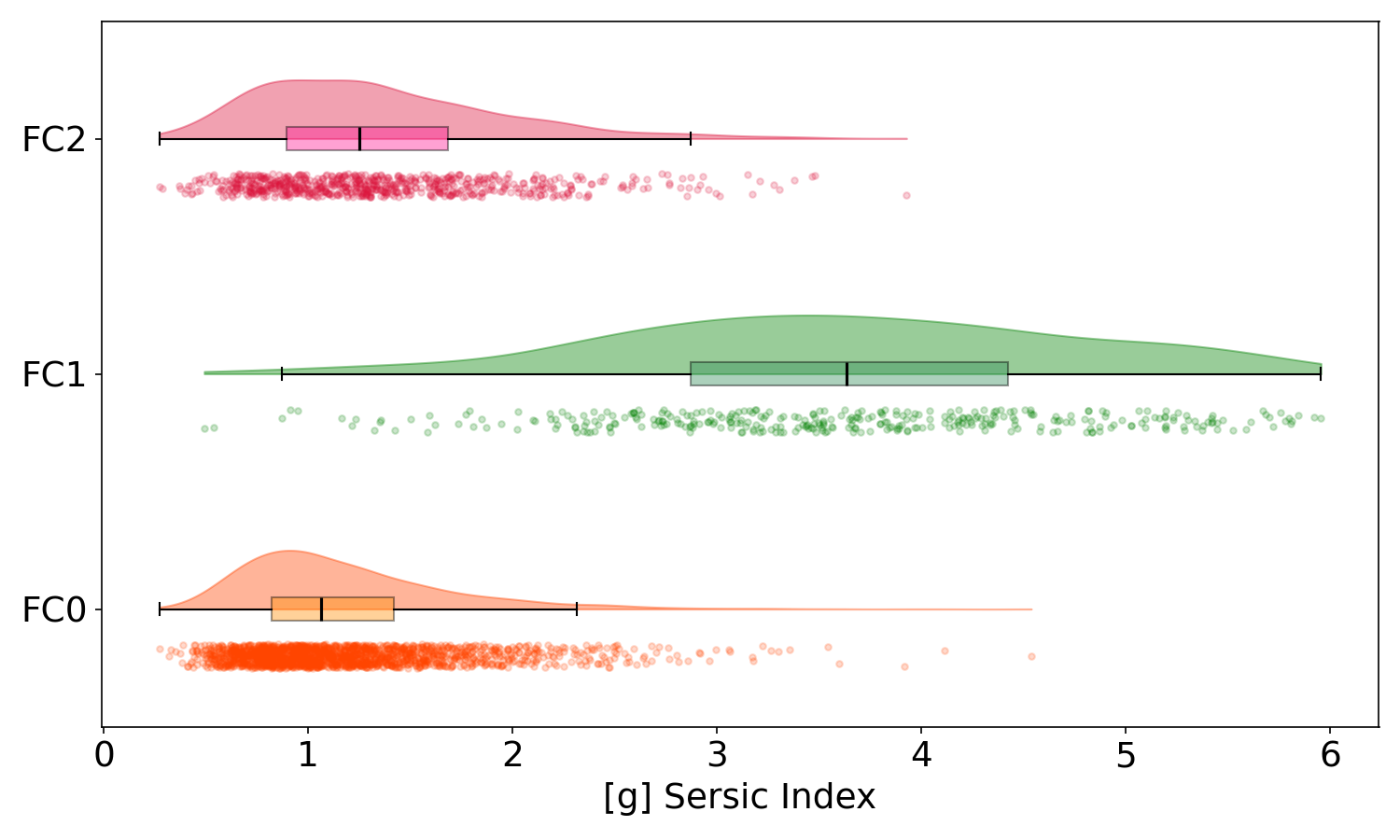}
\end{subfigure}
\hspace{0.5cm}
\begin{subfigure}{0.45\textwidth}
\centering
\includegraphics[width=\linewidth]{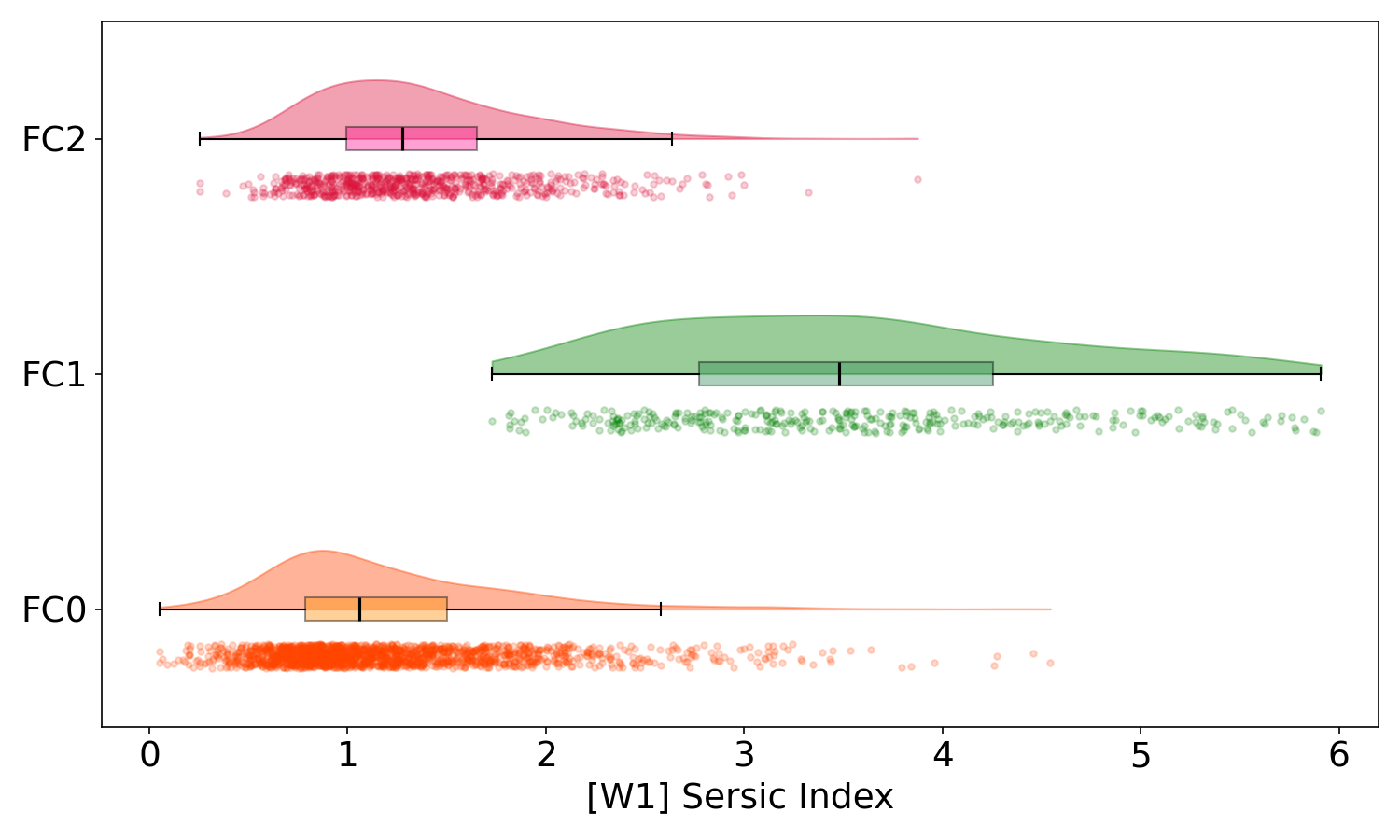}
\end{subfigure}
\hspace{0.5cm}
\begin{subfigure}{0.45\textwidth}
\centering
\includegraphics[width=\linewidth]{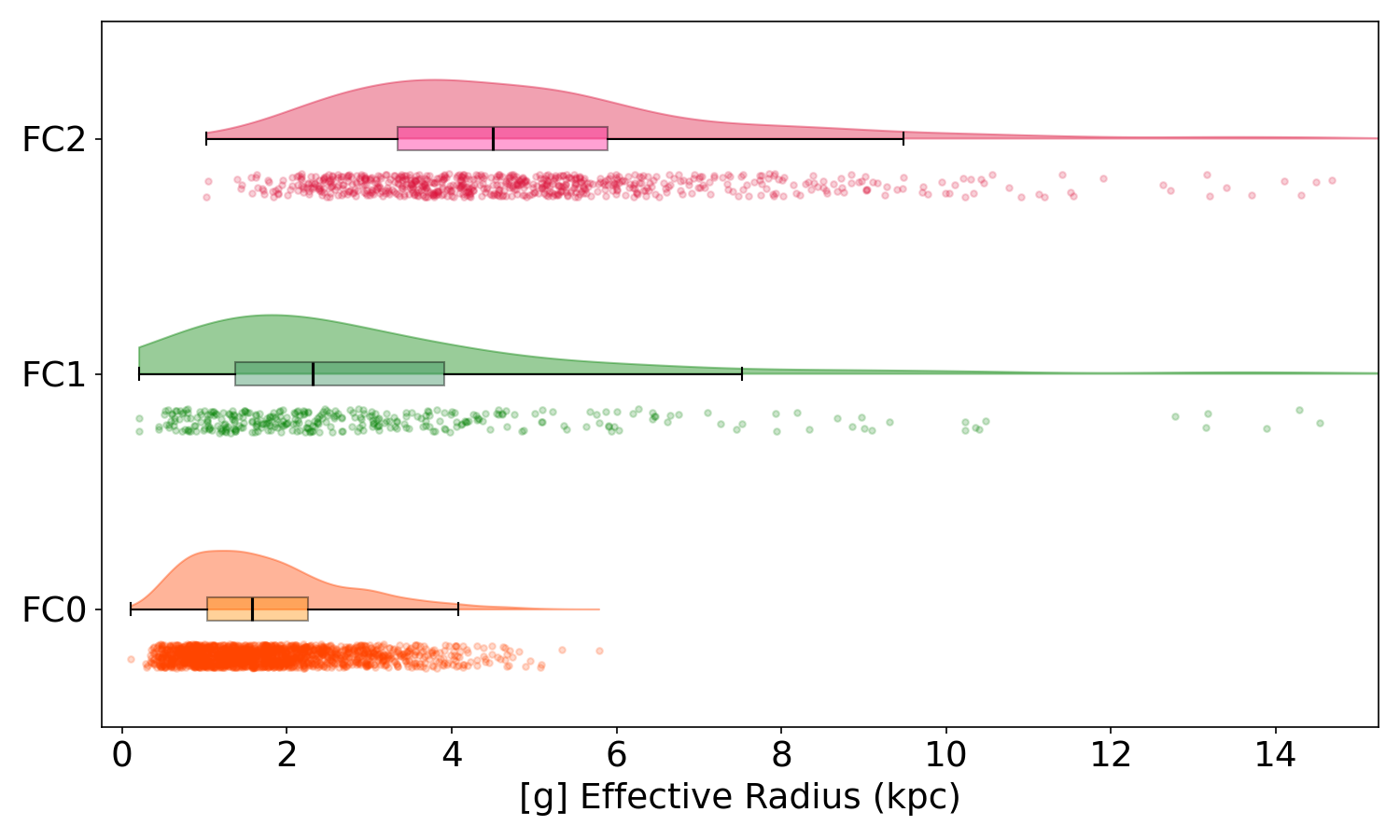}
\end{subfigure}
\hspace{0.5cm}
\begin{subfigure}{0.45\textwidth}
\centering
\includegraphics[width=\linewidth]{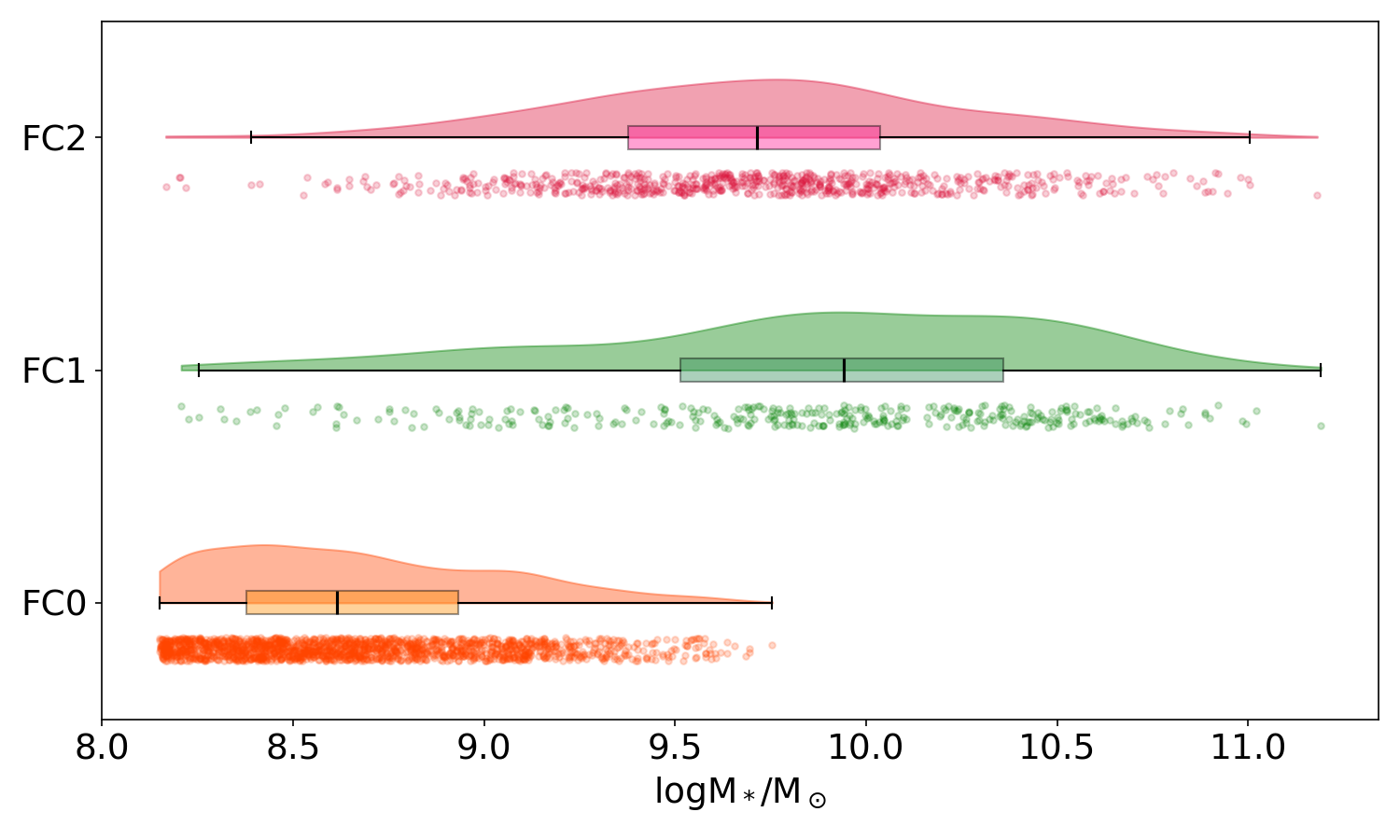}
\end{subfigure}
\hspace{0.5cm}
\begin{subfigure}{0.45\textwidth}
\centering
\includegraphics[width=\linewidth]{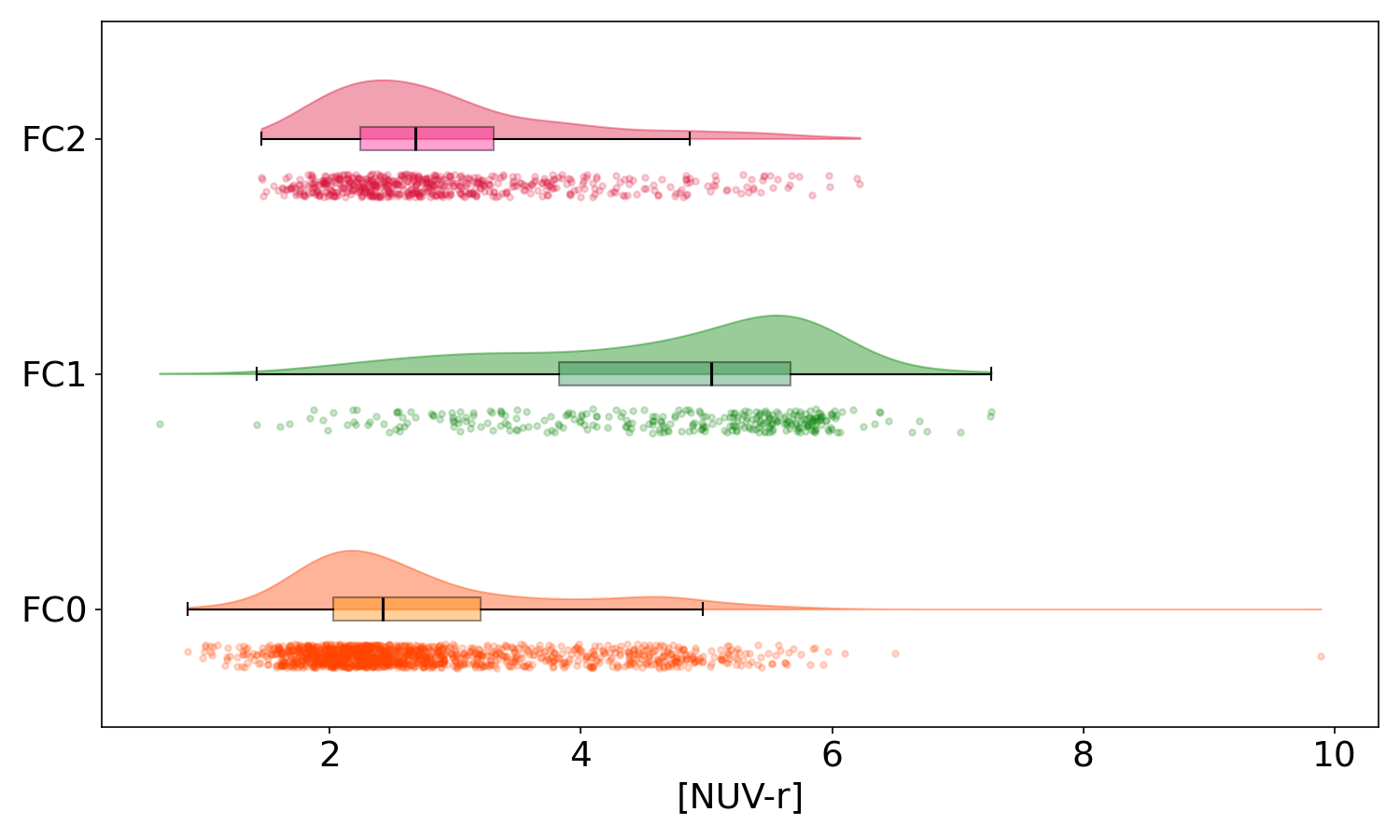}
\end{subfigure}
\hspace{0.5cm}
\begin{subfigure}{0.45\textwidth}
\centering
\includegraphics[width=\linewidth]{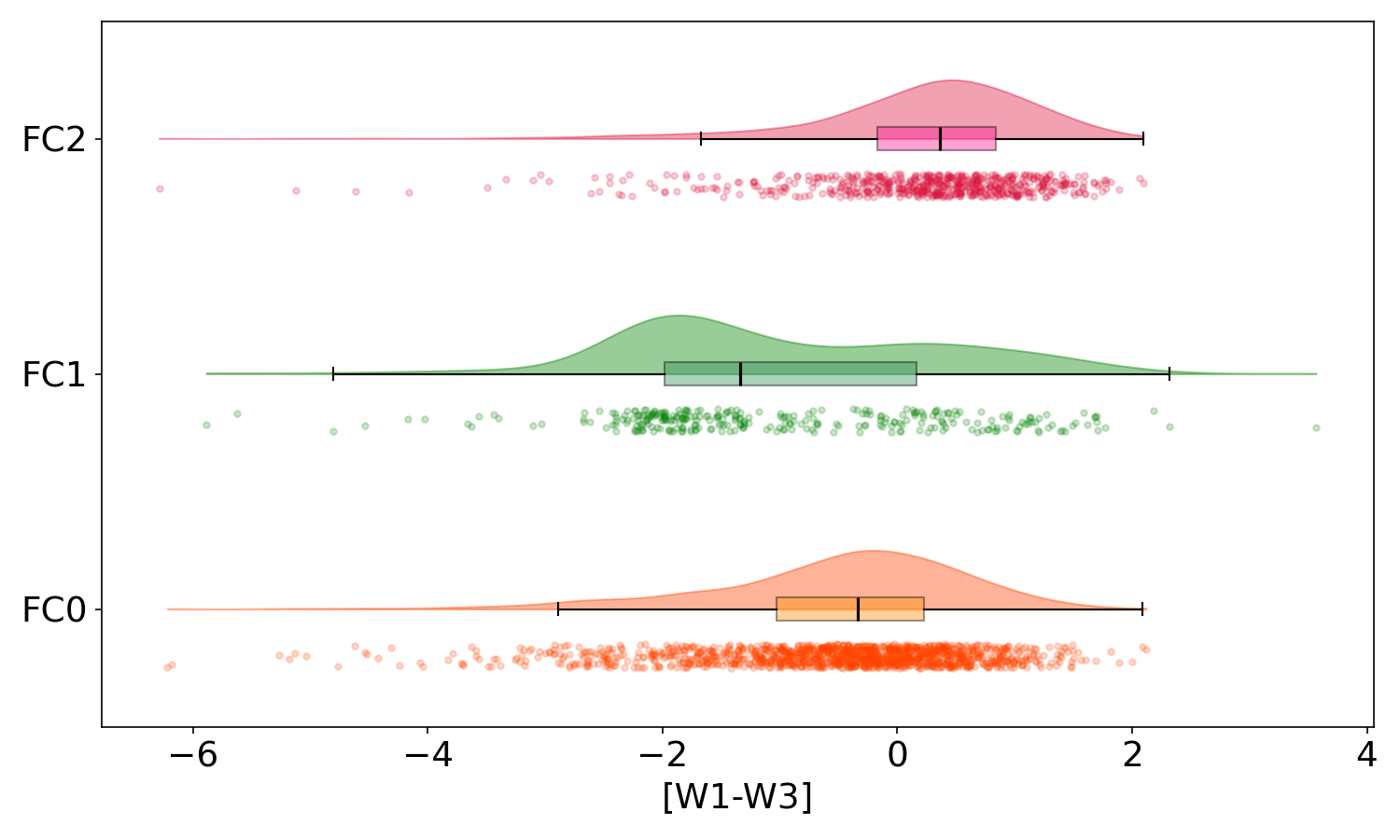}
\end{subfigure}
\caption{Raincloud plots for the \textit{g}-band and W1 \textit{n} (row one), \textit{g}-band and stellar mass (row two), and NUV-r and W1-W3 colors (row three). The shaded curve is the density shape of the data, where the black vertical line is the median and the endpoints are 1.5 times the IQR. The box extends from the 25\% to the 75\% quartile. For clarity, we hide five \textit{g}-band galaxies with $15<R_e<20$ kpc.} 
\label{fig:rainclouds}
\end{figure*}



In Figure \ref{fig:rainclouds}, we visualize the physical results of how k-means clustering partitioned our galaxy sample. The first two rows show the \textit{g}-band R$_e$, stellar mass, and \textit{n} features of each FC -- the four features which our k-means algorithm used to partition the data. FC1 and FC2 span almost the full range of R$_e$ and stellar mass while FC0 is much narrower. There is a similar trend for FC0's \textit{n} distribution, albeit with a longer tail toward larger values. We see comparable patterns in Figure \ref{fig:faux_pca}.

Concentrating on the medians of these distributions, FC0 tends toward low \textit{n} in both bands, small \textit{g}-band R$_e$, and low stellar masses. Despite the wider ranger of \textit{n}, the FC1 medians are clearly shifted toward higher indices, small \textit{g}-band R$_e$, and higher stellar mass. FC2 shares traits with both classes, having low \textit{n}, large R$_e$, and high stellar mass.

Our clustering model does not correspond directly to traditional Hubble Types but does map broadly onto various properties associated with these traditional types. We will henceforth refer to our FCs with this nomenclature:
\begin{itemize}
    \item Dwarf Galaxies (FC0)
    \item Spheroids (FC1)
    \item Extended Disks (FC2)
\end{itemize}

\begin{figure*}[t]
\includegraphics[width=\linewidth]{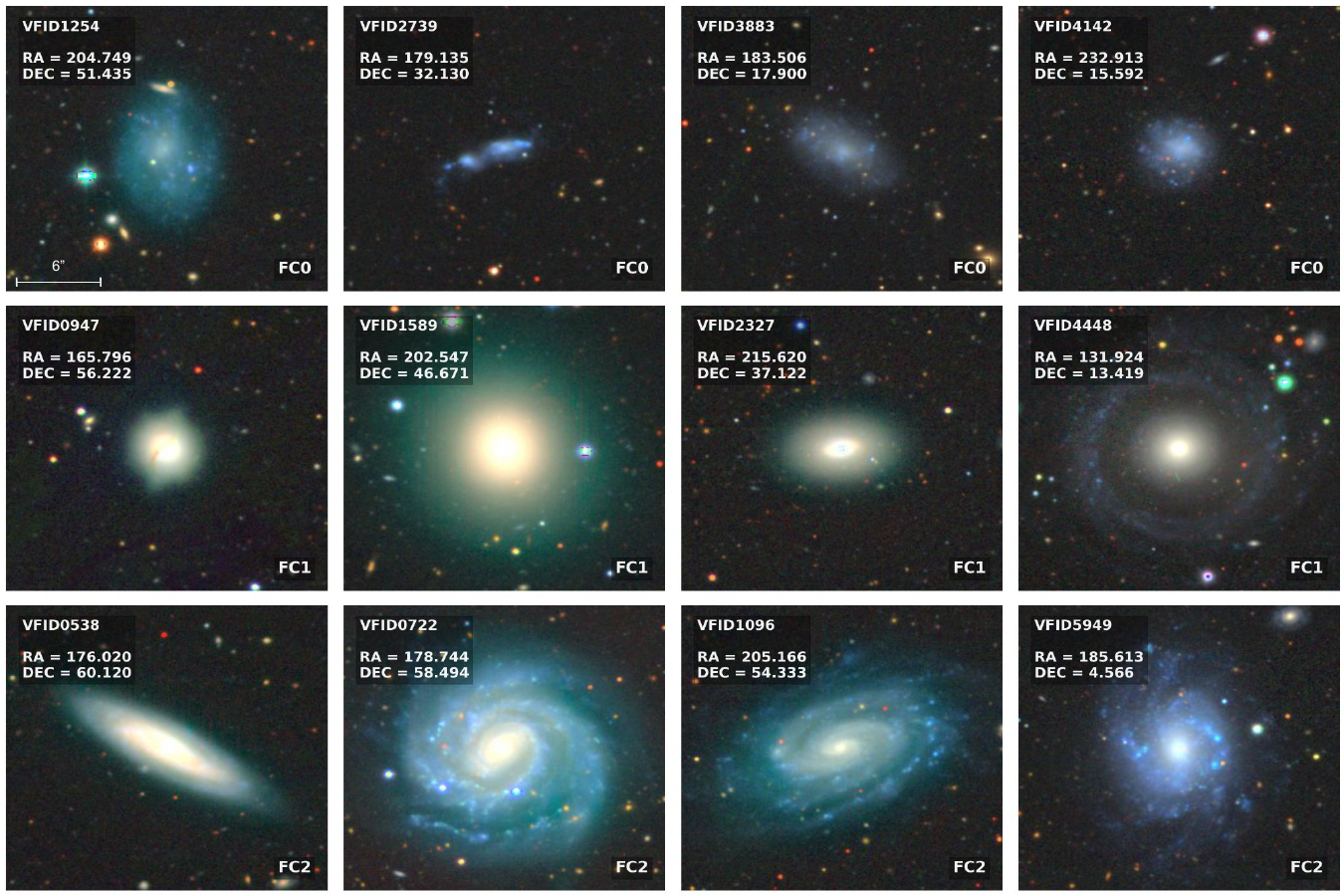}
\centering
\caption{A mosaic of representative galaxies from each FC. The VFID from our Virgo catalog, along with the RA-DEC coordinates of the galaxy's center, are provided in the upper left of each panel. Each row corresponds to an FC, with explicit labels in the bottom right of each panel for reference. The upper leftmost galaxy panel includes a physical scale, which applies to every image in the mosaic.}
\vspace{.5mm}
\label{fig:reps}
\end{figure*}

In Figure \ref{fig:reps}, we show four representative galaxies from each FC, all at the same scale factor of 0\farcs 7 per pixel. The first row includes galaxies from FC0, which largely appear diffuse and dwarf-like. The second row, FC1, have galaxies which are more elliptical with relatively little pronounced structure; and FC2 in the bottom row comprises galaxies with bulges and well-resolved disks. 

The third row of Figure \ref{fig:rainclouds} shows the data for the NUV-r and W1-W3 color magnitudes for our three FCs. The dwarfs and disk populations are shifted toward lower NUV-r, indicating that these galaxies generally are brighter in the wavelength regime associated with active and unobscured star formation. The spheroid population shows a much more extended distribution but peaks at higher NUV-r, signaling older stellar populations with comparatively few young stars. There is a hint of a bimodality in the colors for the spheroids, with the strongest peak at the redder end of the axis. While there may be spheroid galaxies in this structural class actively forming stars, they are small in number compared to the passive spheroids.

The distinction between the three FCs is less stark for W1-W3, but the median color for the spheroids is lower by around one magnitude compared with the disks. We also note a bimodality of the W1-W3 distribution for the spheroids, again suggesting that there are both passive and active spheroids in this class. However, since the strongest peak emerges blue end, the fraction of passive spheroids appears to be more dominant. 


\subsection{Environment Distribution of Feature Classes}

\begin{figure}
\includegraphics[scale=0.33]{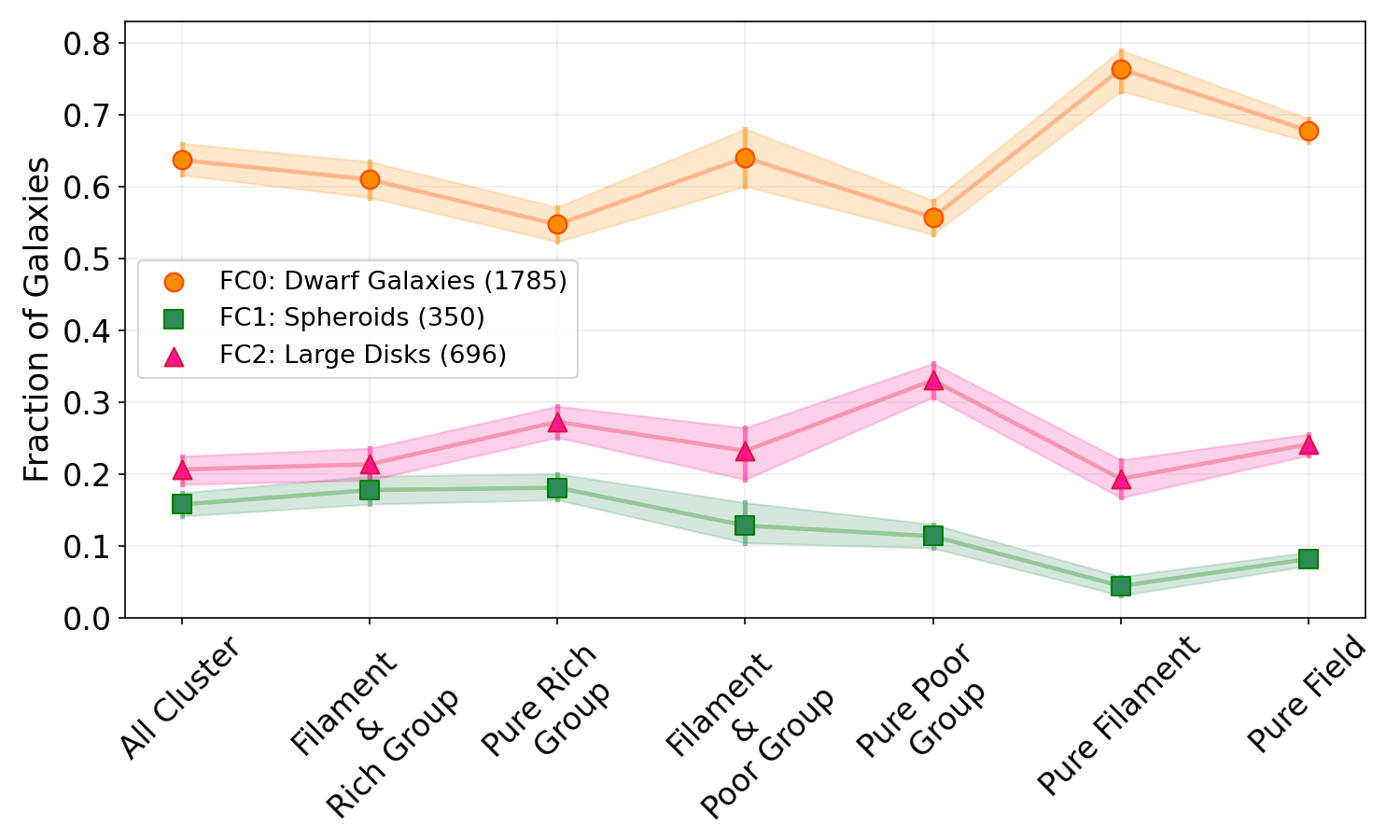}
\centering
\caption{The FC composition of each environment, arranged from most dense to least dense. ``All Cluster" refers to galaxies that are either strictly cluster galaxies as well as those that may belong to both the cluster and rich group, while environments combined with an ``\&" contain galaxies that are in both environments only. The different FCs are marked in the legend, with the total number of galaxies in each given parenthetically. Added vertically, each environment sums to 1. We calculate the 68\% confidence intervals, shown in the shaded region, using bootstrapping on the mean (subset/total).}
\vspace{.5mm}
\label{fig:envfraction}
\end{figure}

Using the environment labels described in Section \ref{sec:env}, we separate galaxies belonging to every environment and determine the fraction of galaxies in these environments belonging to each FC. We show these results in Figure \ref{fig:envfraction}, which plots the fractional subset of each FC for environments arranged from most-to-least dense. The shaded regions represent the 68\% confidence intervals found using bootstrap resampling.

As expected given its abundance of galaxies, the compact dwarf class dominates the fractional composition of every environment. We observe a nearly 10\% increase in dwarf galaxies from the cluster and rich group regimes to the isolated environments, while the large disk class shows consistent membership among the environments. Moving to spheroids, which are not as numerous as the other two FCs, there is a significantly higher fraction of these galaxies in the densest environments compared to their presence in the least dense environments. This almost monotonic dip in spheroid concentration from cluster to field is in accordance with the morphology-density relation, which argues that ETGs are more common in denser environments \citep{dressler1980}. Otherwise, we do not discern any meaningful trends in this figure, suggesting that the relative abundance of different FCs is independent of environment. However, as we expound in Section \ref{sec:ttype}, there is a mix of classical morphologies within our dwarf population that complicates the interpretation of this FC.

\subsection{Star-Formation Rate - Stellar Mass Relation of Feature Classes}
\label{sec:logsfr}


We calculate the star-forming main sequence of our full VFS dataset by first fitting a line to the subset of galaxies with log(sSFR)$>$-11.5. The reason for this limit is because galaxies below this value produce UV and IR emission that is likely dominated by sources not associated with star formation \citep{salim2018}. The resulting equation for our main sequence is
\begin{equation}
    \log(\text{SFR}) = 0.80\log(M_*) - 8.56.
\end{equation}



\begin{figure}[t!]
\centering
\centering
\includegraphics[width=\linewidth]{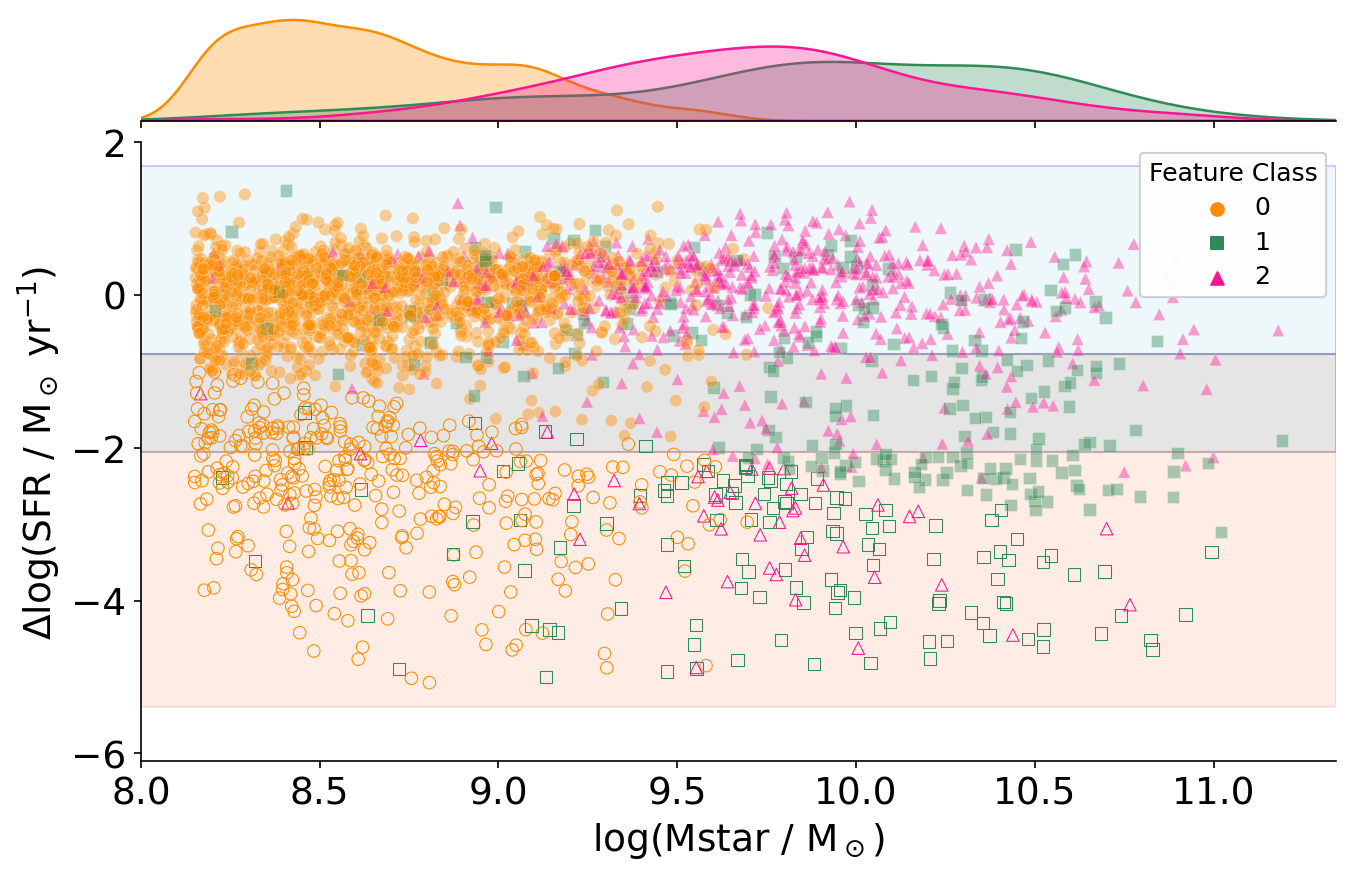}
\caption{Star formation rate main sequence offset ($\Delta$log(SFR)) vs. stellar mass (log(M$_\star$)) for our sample. The points are again color-coded according to their respective FC. $\Delta$log(SFR) is calculated using the main sequence equation in Section \ref{sec:logsfr}. The offset axis is separated into three regimes: main sequence (top), suppressed (middle), and passive (bottom). }
\label{fig:dsfrmstar}
\vspace{.5mm}
\end{figure}

We use this equation to calculate $\Delta$log(SFR), defined as the difference between a galaxy's \texttt{CIGALE} log(SFR) and the log(SFR) estimated by the main sequence. We plot the distribution of $\Delta$log(SFR) versus \texttt{CIGALE} log($M_\ast$) for the three FCs in Figure \ref{fig:dsfrmstar}, along with closed/open shapes depending on whether \texttt{CIGALE} SFRs were above or below the log(SFR) = -3 floor, respectively. We observe an abundance of these objects in the lower left quadrant of the distribution, corresponding to dwarf galaxies with low SFR that nevertheless have a sufficiently high S/N in either the NUV or W3. The kernel density estimation curves (KDEs) on the auxiliary top axis, which are smoothed tracers of the data distribution, show that dwarf galaxies indeed peak at the lowest stellar masses but primarily occupy the main sequence regime of star formation. The extended disks (magenta) and spheroids (green) span the full range of stellar masses for our population.

\begin{figure}[h]
\includegraphics[scale=0.41]{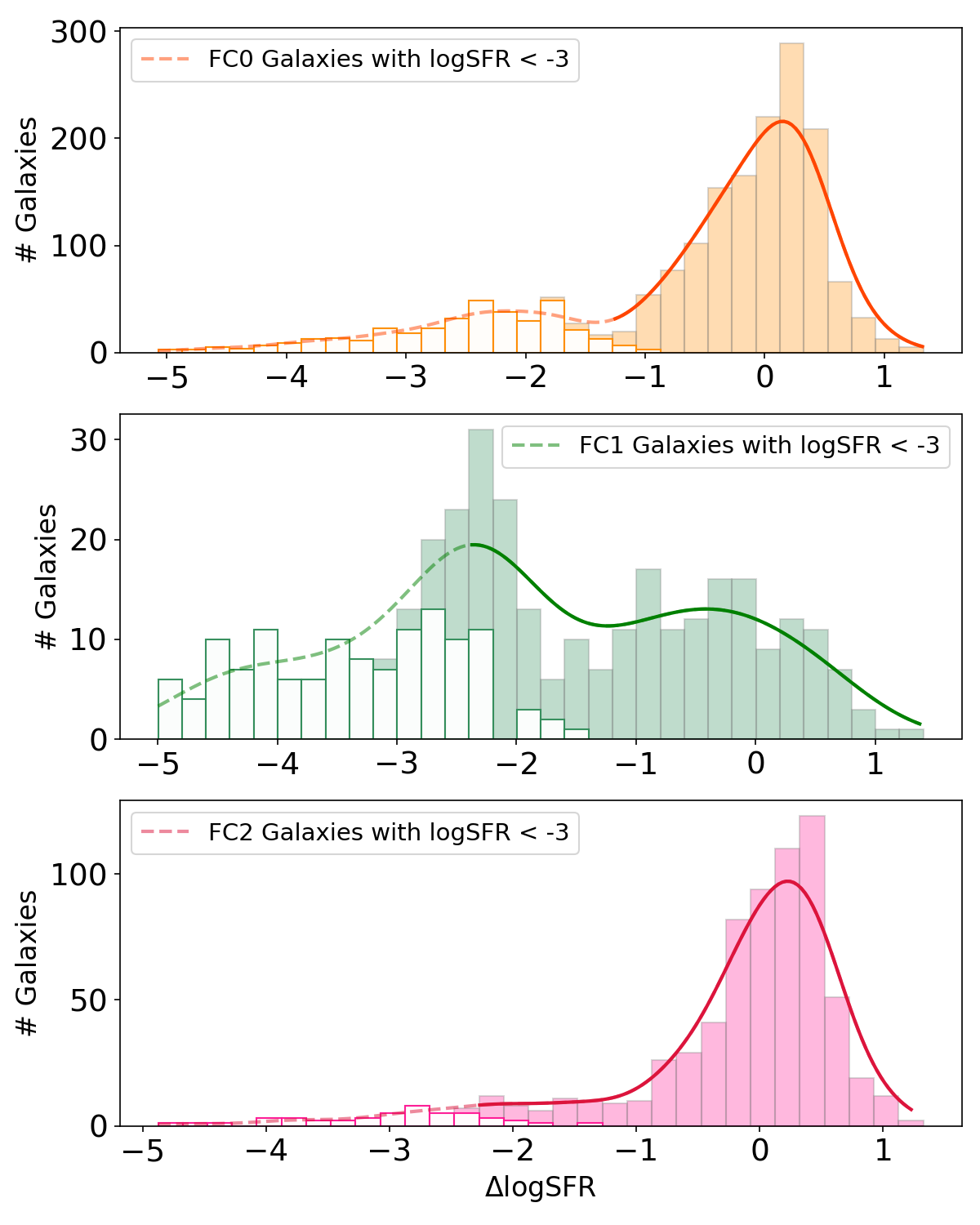}
\centering
\caption{Distribution of main sequence offsets as a function of FC. The y-axis indicates the number of galaxies within each bin of $\Delta$logSFR. The colored bins are the full set of FC galaxies, whereas the white bins are the number of galaxies within the full bin with logSFR $<$ -3. The curves trace the KDEs for the histograms, with the dashed line signaling where the percentage of logSFR $<$ -3 galaxies compared to the total number of binned galaxies is at least 40\%.}
\vspace{.5mm}
\label{fig:kde_dsfr}
\end{figure}

Figure \ref{fig:kde_dsfr} shows the distribution of the $\Delta$log(SFR) for each FC, both in the form of raw data (histograms) and kernel density estimates (KDEs). We explicitly plot with a dashed KDE line the range of $\Delta$log(SFR) where the \texttt{CIGALE} log(SFR) $<$ -3, though we note that applying a log(SFR) = -3 floor does not affect either the behavior or our interpretation of the results. We perform a Kolmogorov-Smirnov (K-S) test on the full $\Delta$log(SFR) range to assess whether the three FCs are statistically distinct, and summarize these results in Table \ref{tab:ks}. None of the distributions are drawn from the same galaxy population, as the p-values for each FC pair demonstrate with a significance $>5\sigma$. The spheroid and extended disk populations are the most distinct in their SF activity with p = $4.35\times10^{-67}$, followed by the spheroids and dwarf galaxies with p = $4.39\times10^{-55}$.

To further analyze the main sequence offsets between FCs, we split the $\Delta$log(SFR) axis into three regimes, represented with shaded rectangles in Figure \ref{fig:dsfrmstar}. We define the main sequence regime to be $\Delta$log(SFR) $\pm$1.5$\sigma$, where $\sigma = 0.51$ dex is the standard deviation of the scatter about the main sequence line for galaxies above log(sSFR) $>$ -11.5. The 1.5 multiple is motivated by analysis done in \citet{finn2023} to encompass a sufficiently large fraction of main sequence galaxies. The middle rectangle highlights what we call the suppressed galaxy population in our sample, ranging from -4$\sigma <$ $\Delta$log(SFR) $< -1.5\sigma$. Any galaxies below the -4$\sigma$ limit are within the passive regime and thus forming few to no stars. We then calculate the fraction of galaxies in a given FC that belong to one of the three regimes. While our fractions use the formal SFR values, we again find that applying a log(SFR) = -3 floor does not affect either the behavior or our interpretation of the results. 

We find that most of the dwarf (69\%) and extended disk (78\%) galaxies lie near the main sequence and relatively few are classified as passive (0.19\% and 0.10\%, respectively). The opposite holds for the spheroids, which are most prominent in the passive regime (60\%) and sparser around the main sequence (25\%). We observe no significant difference between the FC fractions in the suppressed regime (10-18\%), suggesting a mix of morphologies and SFRs. This finding is in line with work exploring the role of different quenching pathways in the diversity of morphologies in the green valley \citep[e.g.][]{schawinski2014, vulcani2015}. We therefore quantitatively confirm that, overall, our spheroid population is predominately underneath the SF main sequence, while the bulk of the dwarf and extended disk galaxies lie within 1.5$\sigma$ of the main sequence. A fraction of the dwarf galaxies with the lowest stellar masses tend to have relatively lower SFRs despite being structurally equivalent under k-means clustering, but these galaxies do not represent more than 30\% of that FC. We explore this further in Section \ref{sec:ttype}.

\begin{table}
\centering
\begin{tabular}{|c|c|c|}
\hline
FC Pair &  $\Delta$logSFR p-value \\
\hline
Dwarf Galaxies \& Spheroids  & 4.93$\times 10^{-55}$ \\
Dwarf Galaxies \& Extended Disks  & 1.10$\times 10^{-10}$ \\
Spheroids \& Extended Disks  & 4.35$\times 10^{-67}$ \\
\hline
\end{tabular}
\caption{K-S test p-values comparing $\Delta$logSFR between FC pairs. The p-value indicates the statistical significance of the two FCs being distinct populations.}
\label{tab:ks}
\end{table}

\subsection{Suppression of SF in Galaxies of Different Structural Types}
\label{sec:cumulative}

\begin{figure*}
\begin{subfigure}{0.49\textwidth}
\centering
\includegraphics[width=\linewidth]{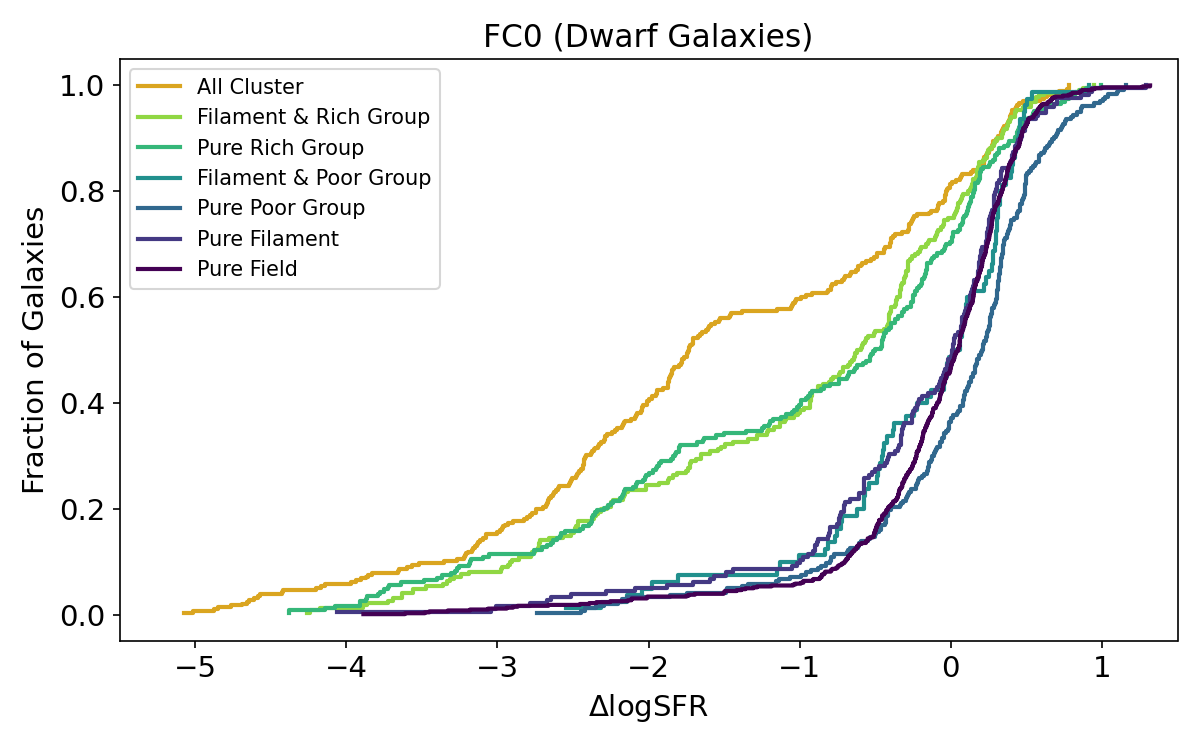}
\end{subfigure}
\hfill
\begin{subfigure}{0.49\textwidth}
\centering
\includegraphics[width=\linewidth]{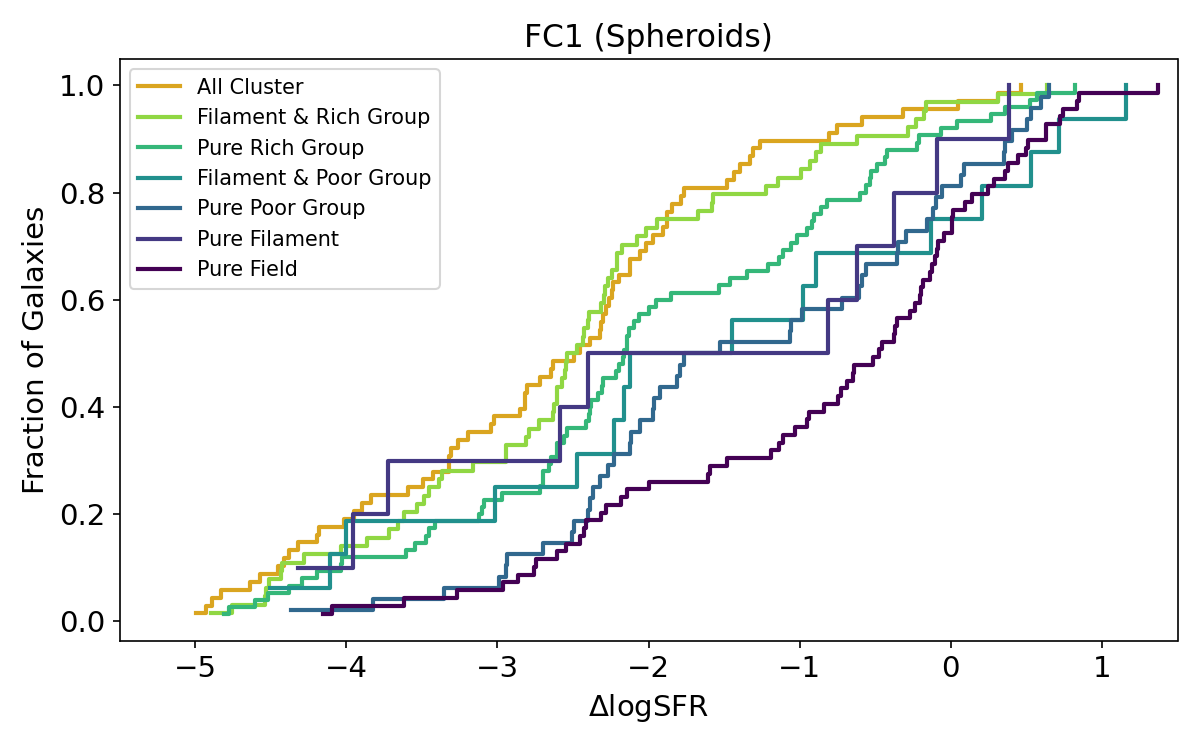}
\end{subfigure}
\hfill
\begin{subfigure}{0.49\textwidth}
\centering
\includegraphics[width=\linewidth]{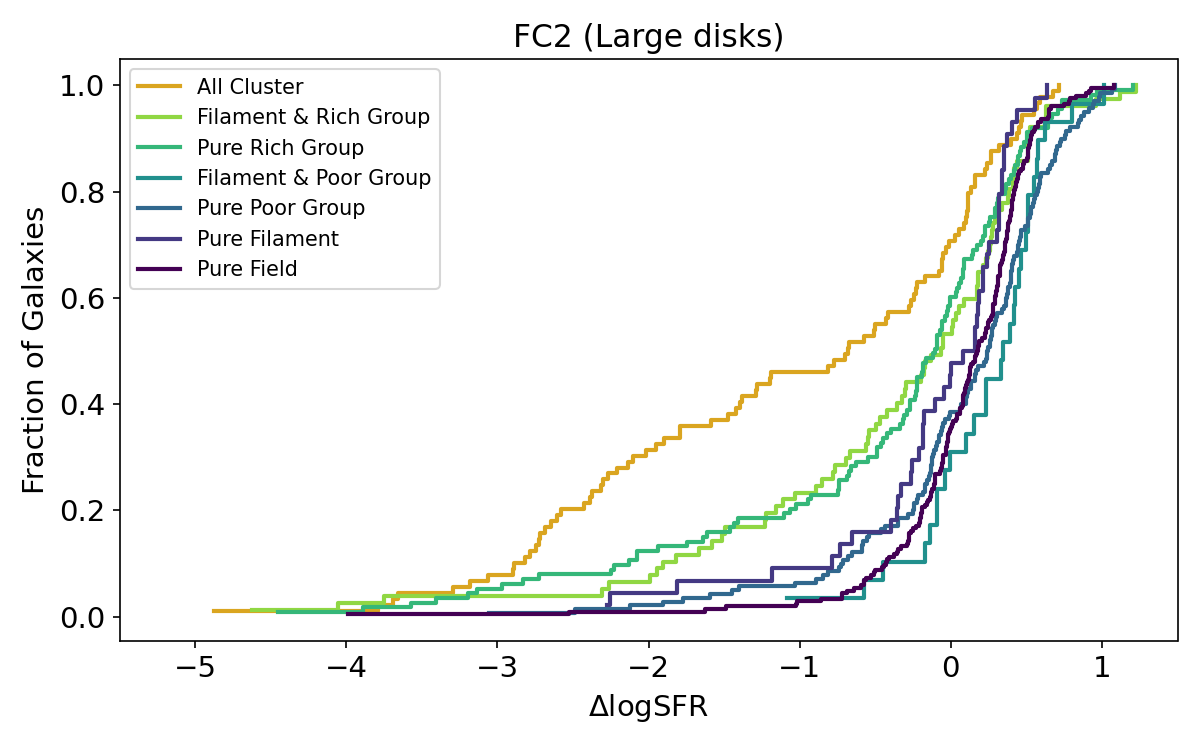}
\end{subfigure}
\caption{Cumulative histogram distributions of $\Delta$logSFR for dwarf galaxies (top left), spheroids (top right), and extended disks (bottom center). Each color traces the distribution of galaxies belonging to the environment indicated in the legend. The curves span the full range of log(SFR), though the use of the log(SFR) = -3 floor does not affect the validity of the results.} 
\label{fig:cum_hist}
\end{figure*}

Figure \ref{fig:cum_hist} shows the cumulative histograms of $\Delta$logSFR for each of our three FCs. Every line corresponds to the galaxies in a particular environment, with the lighter colors representing denser environments. If a line is shifted toward the right, then the galaxies in that subpopulation tend to be more star forming, while lines to the left are more suppressed. In all three FCs, galaxies belonging to the Virgo cluster environment show the lowest SFRs. We also see that the lowest density environments harbor the most star-forming galaxies in each FC. 

The line distributions then become class-dependent. For the dwarf galaxies and large disks, there is no strong environmental effect on SFR until we reach the densest environments, at which point there is a sharp drop to lower SFRs. The overall suppression effects of the densest environments are stronger for dwarf galaxies, as the median of the densest environment curves are shifted more toward lower SFRs compared with the larger disk galaxies. We also see more suppression overall of the dwarf galaxies in the rich group and cluster environments compared to the large disks.

For spheroidal galaxies, there is a systematic shift toward lower SFRs for all environments, including the field, compared to the other FCs. A more unexpected observation, however, is that there is a smoother progression to lower SFRs from field to cluster for the spheroidal galaxies compared to the dwarf and extended disk populations.

To determine if these trends are due to the morphology density relation changing the mix of morphologies within every FC, we test whether these trends are due to variations of \textit{n} within each FC. We first check for global correlations between \textit{n} and $\Delta$log(SFR). For both W1 and \textit{g}-band, we compare the median \textit{n} for galaxies with $-0.5 < $ $\Delta$log(SFR) $< 0.5$ and those with $\Delta$log(SFR) $> -0.5$. We find that for every FC, the difference in between the lower and higher $\Delta$log(SFR) galaxies are consistent within the 68\% confidence intervals on the medians. That is, the suppression of star formation within the FCs is not accompanied by changes in morphology.

We then compare the \textit{n} distributions in each environment to determine if the $\Delta$log(SFR) changes correspond with changes in \textit{n}. If this were the case, we would expect galaxies to move to higher \textit{n} as environment density increases. Instead, we observe that the \textit{n} distributions between any two environments are either statistically indistinguishable as determined by a K-S test; or that the differences in their their median \textit{n}, $<0.4$, which is not a large enough separation to account for the differences we find in $\Delta$log(SFR).

\section{Discussion}
\label{sec:discussion}

In this section we explore the implications of our UML-derived FCs. We omit the logSFR = -3 floor for all plots but verify that the statistical significance of our results does not change with this choice.


\subsection{Environment Quenching Effects per FC: Comparison with Theory and Observation}
\label{sec:theory}

In Section \ref{sec:cumulative}, we observed trends in the distribution of $\Delta$log(SFR) across our three FCs. Spheroidal galaxies follow a smoother progression to lower SFRs with increasing environment density, whereas dwarf and large disk galaxies primarily show environmental shifts toward lower $\Delta$log(SFR) in rich groups and the cluster. While the trends for dwarfs and large disk classes are qualitatively similar, the impacts of rich group and cluster environments on the dwarf population are more pronounced. In this section, we will discuss possible interpretations for these different suppression trends for our three FCs motivated by existing work, beginning with dwarf galaxies. We also demonstrate that, in regards to the large disk and spheroid classes, any such interpretations should consider not only stellar mass but also the broader structural parameters of the galaxies.

The sensitivity of dwarf galaxies to their host environment is well established in the literature, both in observations and simulations \citep[see e.g.,][]{gavazzi2010, peng2010_lowmass, geha2012, fillingham2015, wetzel2015, xie2020, donnari2021, romero-gomez2024}. From one perspective, the shallower gravitational potentials make dwarf galaxies more susceptible to hydrodynamical outer disk stripping and tidal perturbations. At the same time, dense environments prevent the accretion of new material onto low-mass galaxies while high-mass systems can more easily retain their reservoir against this depletion \citep{hahn2007, lee2018, musso2018, zakharova2026}. Taken together, these factors explain why the rich group and cluster environments produce a stronger shift toward lower $\Delta$log(SFR) in our dwarf galaxy population compared with their large disk counterparts. 

Recent simulation studies have also shown that the evolution of low-mass galaxies is mainly driven by the group and cluster environments \citep[e.g.][]{xie2020, donnari2021, zakharova2026}. This result is indeed consistent with our dwarf population. If we observe how the low-mass and high-mass ends of our large disk and spheroid galaxies compare with these predictions, however, we find that stellar mass may not be enough to predict the suppression efficiency of a galaxy's host environment. 

Starting with the large disk galaxies, we have seen that the patterns of $\Delta$log(SFR) suppression for this broad class are comparable to those of the dwarf class. If we split the large disks into log(M$_\ast$) $<9$ and log(M$_\ast$) $\geq9$ subclasses, the distribution of high-mass galaxies still follows this same trend of being principally affected by rich group and cluster environment. Similarly, for the spheroidal galaxies, the gradual nature of environmental quenching efficiencies we see in \ref{fig:cum_hist} holds regardless of whether we plot all galaxies in FC1 or only those above or below a certain stellar mass threshold, such as the log(M$_\ast$) $<9$ cut we chose for the large disks. Stated differently, stellar mass alone does not account for the observed environment suppression patterns in our three FCs, and a galaxy's overall structure plays a role in determining whether environmental quenching becomes significant only in the richest global environments or proceeds gradually with density. 

To reiterate what we observe for our spheroid class, these galaxies are more suppressed in all environments galaxies than the other two FCs. Furthermore, they display a more gradual progression as environmental density increases. One possible interpretation is that these galaxies had already undergone in situ quenching related to their structural morphology. These intrinsic processes range from stellar and AGN feedback \citep[e.g.][]{croton2006, lopez2014, schawinski2014, suin2024} to the build up of the galactic bulge \citep[e.g., morphological quenching;][]{euclid2026, lin2026, polletta2026}. Morphological quenching refers to the suppression of star formation because a galaxy’s structure makes its gas disk gravitationally stable, rather than because the galaxy has lost its gas due to environmental influences. This idea is particularly compelling because we observe high-\textit{n} galaxies with low SFRs in the lowest density environments, suggesting that members of this spheroidal class had an extensive intrinsic evolution prior to experiencing the processing of their current host environments. 

\subsection{Linking Structural Types and Traditional Morphologies}
\label{sec:ttype}

Using k-means we partitioned our subset of galaxies in and around the Virgo cluster into three FCs, which the input feature distributions revealed contain predominantly dwarfs (FC0), spheroids (FC1), or extended disks (FC2). Although we can broadly label these classes with names that have morphological connotations, it is unclear how our objective classification technique compares with a more subjective morphology classification scheme. Establishing this connection is the goal of our current discussion.

\begin{figure}[h]
\includegraphics[scale=0.335]{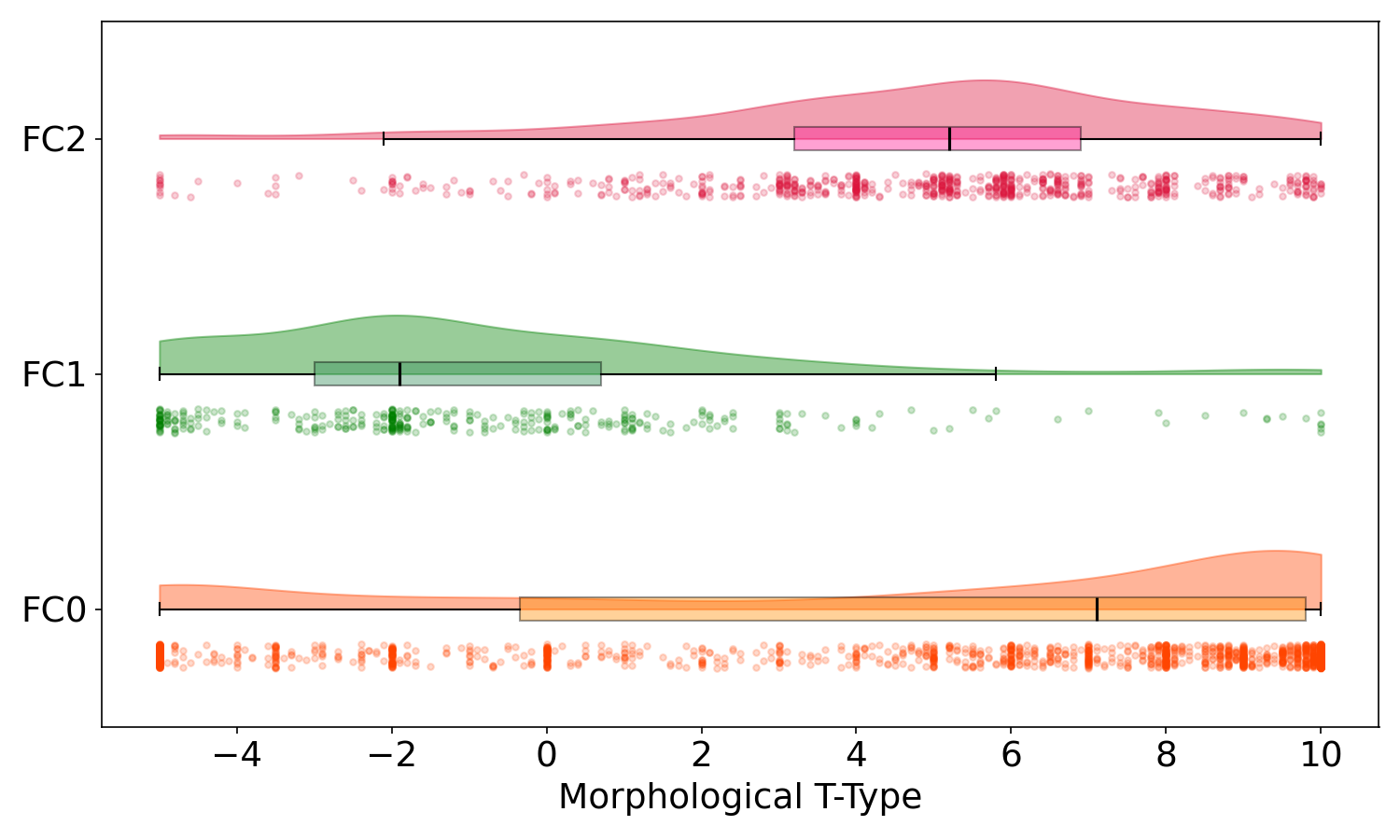}
\centering
\caption{Raincloud plots showing the distribution of morphological T-types for our three FCs: dwarf galaxies (FC0), spheroids (FC1), and large disks (FC2).}
\vspace{.5mm}
\label{fig:ttype}
\end{figure}

To begin, we take T-type labels from the Virgo galaxy catalog source-matched to the HyperLEDA database \citep{makarov2014, castignanivirgo1}. T-types loosely trace the Hubble tuning fork, with lower values corresponding to early-type galaxies and higher values describing late-type galaxies. We measure the likeness of each FC grouping to these T-type labels in Figure \ref{fig:ttype}. 

We see in the figure that large disks peak at $T\approx5$, spheroids at $T\approx-2$, and dwarfs at $T\approx7$. Looking at the dwarf galaxies specifically, however, we notice a nontrivial population of dwarf galaxies extending to negative T-types. Indeed, of the 1785 galaxies comprising the dwarf class, nearly 30\% have $T\leq0$. 

\begin{figure*}[t]
\includegraphics[scale=0.71]{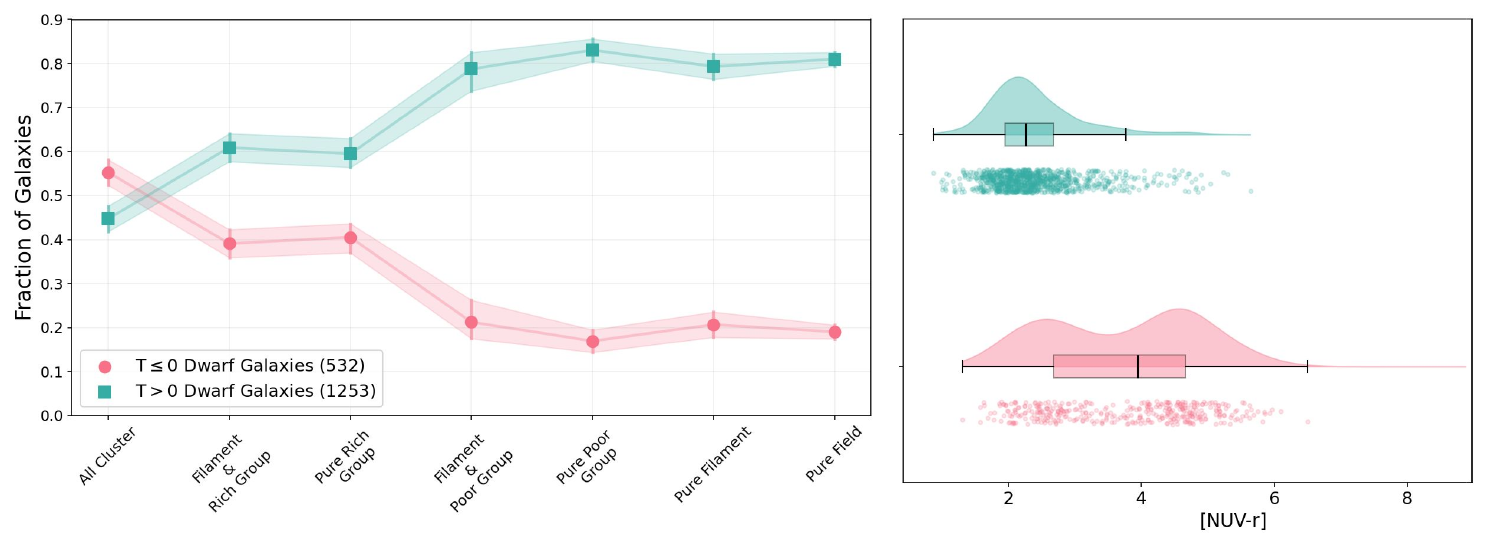}
\centering
\caption{Comparisons of the two dwarf galaxy subpopulations, separated by T-type. The marker shape and color assigned to either T$\leq 0$ (pink circles) or T$> 0$ (teal squares) dwarf galaxies is given in the legend, along with the number of members belonging to each subpopulation. (\textbf{Left}) Distribution of each population according to host environment. (\textbf{Right}) Raincloud plot of the \textit{NUV}-\textit{r} colors.}
\vspace{.5mm}
\label{fig:ttype_comp}
\end{figure*} 

The majority of dwarf galaxies have $T>0$, so the trends of their environment membership, SFR, and NUV-r color dominate the full class. And yet, when we compare these observables between the $T\leq0$ and $T>0$ subpopulations in the dwarf class, the trends are noticeably different \textemdash despite the structural parameters being nearly identical. We plot the subpopulations in Figure \ref{fig:ttype_comp} and find that the $T\leq0$ galaxies show redder NUV-r colors and are more common in denser environments. The $T>0$ galaxies match the general dwarf class trends for all but the environment fractions, as the galaxies almost monotonically climb from 50\% membership of the Virgo cluster to 80\% of the pure field. The $T\leq0$ galaxies, in contrast, are 50\% of the Virgo cluster and around 20\% of the pure field. 


T-type encodes star formation activity since the classification may also include information based on color, spiral arm strength, and clumpiness; so we expect there to be correlations between environment, $\Delta$log(SFR), and T-type. Indeed, a random-forest regression trained on the same structural parameters used in the k-means analysis predicts T-type more accurately when $\Delta\log(\mathrm{SFR})$ is included, increasing the correlation coefficient from $R=0.57$ to $R=0.71$. This, as well as the presence of subpopulations within each FC, suggests that the structural similarity of galaxies does not necessarily imply evolutionary similarity, especially for dwarf galaxies. Likewise, T-type labels are not necessarily indicators of a galaxy's underlying structural parameters.


We want to highlight a caveat in the reported T-types, again in particular to dwarf galaxies. We visually inspected all FC0 galaxies with $T<0$ and found many inconsistencies: multiple galaxies with asymmetrical shapes and non-smooth outskirts, primarily around the main sequence, are marked in HyperLEDA as being lenticular or elliptical. The disconnect may be partially due to the fact that the HyperLEDA catalog consolidates T-types from many sources, and thus from many research groups with their own ideas of morphology classifications. Another complication involves galaxies associated with the spiral galaxies in the Virgo cluster more specifically, where their global star formation is lower than their isolated counterparts but with normal or enhanced SFR in their inner disks. For such galaxies, T-type classifications can be misleading for environmentally-altered morphologies \citep[e.g.][]{Koopmann1998}.

We also note that there is a seemingly artificial build-up of galaxies at $\rm T=-5$ and $\rm T=10$, which are the endpoints of this T-type scale. We find that of our 2831 galaxy sample, 657 (23\%) lie at one of these bounds, with 607 (21\%) being dwarf galaxies. The model also agrees least with the reported HyperLEDA T-types at these endpoints. When galaxies with T = -5 and T = 10 are removed, the R correlation coefficient from the random forest regression model increases to 0.64 for the four structural parameters and to 0.80 when SFR is included. That is, the model's precision improves systematically after we remove these endpoint galaxies.



Additional work is necessary to explore whether k-means clustering algorithms can discern these subpopulations. One approach for this purpose is to recognize that the limitation is not necessarily the algorithm, but the selection of structural parameters. For example, if robust measures of symmetry or velocity dispersion lie orthogonal to R$_e$ and \textit{n} in multidimensional feature space, then their inclusion could possibly reveal finer groupings \citep[e.g., see][]{weijmans2014}. Secondly, a larger subset of galaxies may introduce more partitioning opportunities: more galaxies could mean a feature space with more peaks. Our future work will encompass an order of magnitude more galaxies in the local Universe (Conger et al (in prep.)), allowing us to better assess the robustness of our feature choices for parsing structural diversity using k-means.


\subsection{Star Formation Trends for Central vs. Satellite Galaxies}
\label{sec:censat}

In addition to separating FCs according to the environment classes from \citet{castignani2022}, we use the central and satellite galaxy designations from \citet{tempel2017} as an alternate method to characterize environment. The goal of this analysis is to examine whether a galaxy's structural type correlates with its placement in a group, as well as whether there are systematic differences between central and satellite galaxies within each FC. 

The data in this group catalog is based on SDSS data-release 12 \citep[DR12,][]{alam2015}, and uses a friends-of-friends algorithm to group galaxy neighbors within a certain ``linking length" which varies according to the central galaxy's redshift \citep{tempel2012, tempel2017}. Galaxies within this linking length are considered part of the group system. We match the coordinates of the galaxies in our 2831 subset to this catalog and find matches for 2221 (78\%) galaxies. Of this number, there are 1021 central and 1200 satellite galaxies, with a breakdown of composition for each FC given in Table \ref{tab:censat}. Of the galaxies matched to the \citet{tempel2017} catalog, extended disks tend to be central galaxies whereas spheroids and dwarf galaxies are both more commonly satellites. 

\begin{table}
\centering
\begin{tabular}{|c|c|c|}
\hline
FC & Central Fraction & Satellite Fraction \\
\hline
Dwarf Galaxies  & 0.40 & 0.60 \\
Spheroids  & 0.45 & 0.55 \\
Extended Disks  & 0.60 & 0.40 \\
\hline
\end{tabular}
\caption{Fractional decompositions of each FC into satellite and central galaxies, using the galaxies that were matched to the \citet{tempel2017} group catalog.}
\label{tab:censat}
\end{table}

We look at the distributions of $\Delta$log(SFR) for our three FCs, comparing the satellite and central galaxies. The satellite subpopulations for every class are shifted toward lower SFRs compared to their central counterparts, with a statistical significance of $>3\sigma$ for the spheroids and $>5\sigma$ for the remaining classes. The offset of the satellite galaxies' median $\Delta$log(SFR) compared to that of the central galaxies, for every FC, is -0.5 dex. As with Section \ref{sec:cumulative}, we confirm that these differences in SFR do not coincide with differences in S\'ersic index for both W1 and \textit{g}-band. 

\begin{figure*}[h!]

\begin{subfigure}{1\textwidth}
\centering
\includegraphics[width=\linewidth]{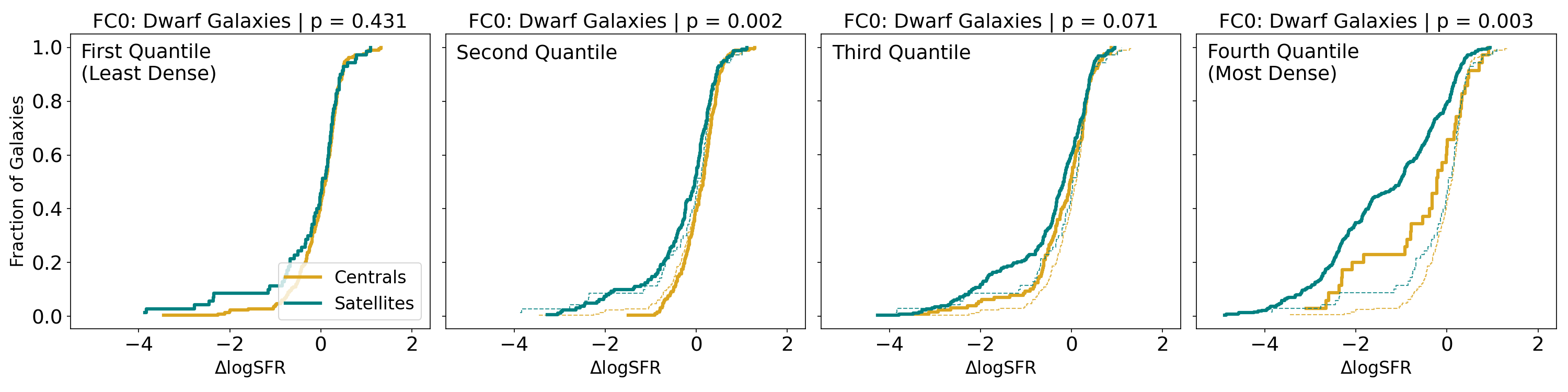}
\end{subfigure}

\vspace{0.01\textwidth}
\begin{subfigure}{1\textwidth}
\centering
\includegraphics[width=\linewidth]{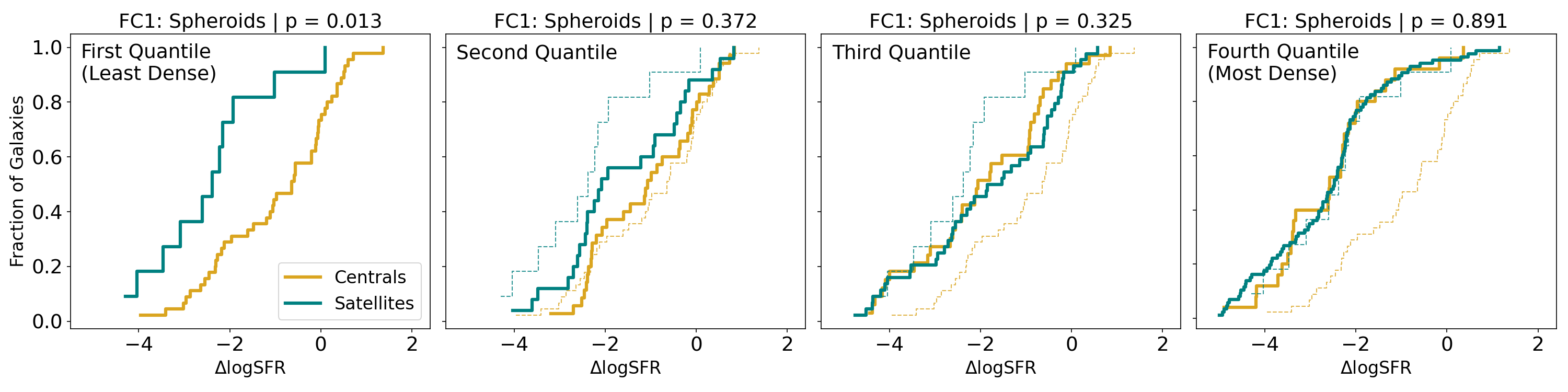}
\end{subfigure}

\vspace{0.01\textwidth}
\begin{subfigure}{1\textwidth}
\centering
\includegraphics[width=\linewidth]{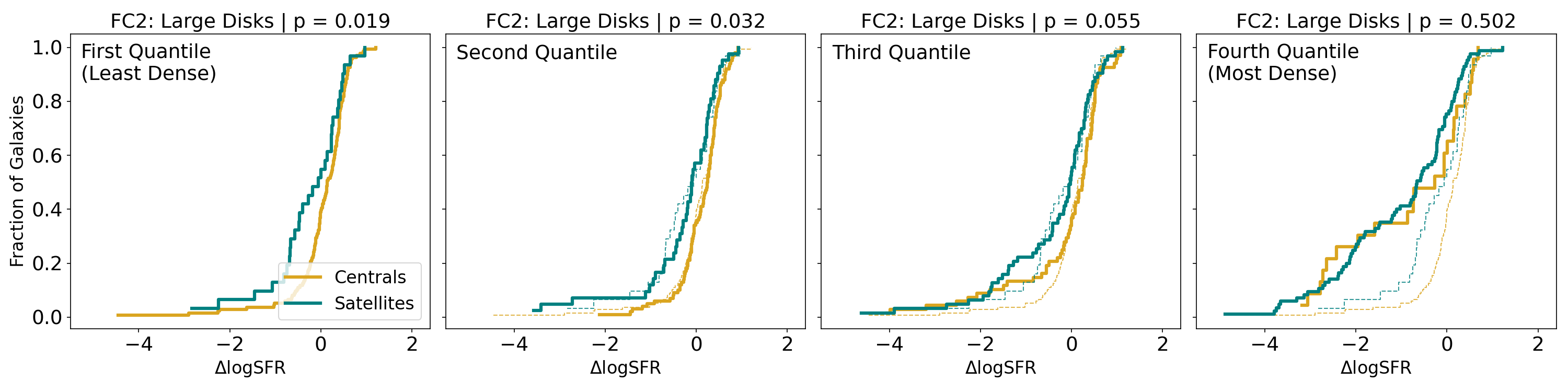}
\end{subfigure}

\caption{Cumulative histogram distributions of main sequence offset for the \citet{tempel2017} central and satellite galaxies in each FC, separated according to local density quantiles. These measures of local density were calculated using the 3D fifth nearest neighbor prescription from \citet{castignani2022}. The title of each panel indicates the FC and the K-S p-value describing the statistical significance of the separation between the two curves. Panels to the right of the first column are shown the first quartile local density histograms as dashed lines, for reference.} 
\label{fig:hist_satcen}
\end{figure*}

Next, we split each FC according local density calculated using the galaxy's fifth nearest neighbor in 3D supergalactic coordinates, the details of which are given in \citet[][]{castignani2022}. We separate our galaxies into four quantiles based on this measurement, arranged from least (first quantile) to most (fourth quantile) dense. In Figure \ref{fig:hist_satcen}, we then plot the cumulative histogram distributions of $\Delta$log(SFR) for central and satellite galaxies within each quantile, again verifying that there are no correlations between the $\Delta$log(SFR) trends and S\'ersic index. We find that for spheroid and large disk galaxies, there are no significant differences between satellites and centrals for any local density bin. For dwarf galaxies, only the second and fourth quartiles show lower SFR for satellites with a $>3\sigma$ significance. We acknowledge that there are relatively fewer galaxies in the spheroid and large disk FCs, so finely binning these classes results in small number statistics that may mask any trends. A larger galaxy sample overall would enable us to test whether small number statistics are concealing trends for these classes. 

We also see indications of systematic suppression of SFRs for galaxies in all FCs in the densest quantile, regardless of whether they are identified by \citet{tempel2017} to be a central or satellite galaxy. For dwarf galaxies, the difference in the median $\Delta$log(SFR) for the first and fourth quantiles is -0.32 dex for centrals and -1.06 dex for satellites. For spheroid galaxies, the difference in medians is -1.93 dex for centrals and -0.09 dex for satellites; and for large disk galaxies we calculate -0.41 dex for centrals and -0.60 dex for satellites. These results suggest that, regardless of a galaxy's structural type, dense global environments suppress \citet{tempel2017} satellites and centrals alike.

\section{Summary}
\label{sec:conclusion}

To isolate the effects of a galaxy's global environment on its evolutionary history, it is imperative to compare galaxies which are structurally similar. To limit the subjective decisions inherent in defining what structurally similar means, UML clustering algorithms provide consistent and reproducible partitions in a multidimensional space for a given dataset. This paper examined the use of our k-means clustering pipeline to partition 2831 nearby galaxies in the extended regions around the Virgo cluster, for which we have full environment characterizations, in structural parameter space. These parameters include \textit{g}-band size, the distribution of W1 and \textit{g}-band light, and stellar mass. We then address how the SFR of our galaxies depends on global environment after controlling for this quantitative morphology. We summarize our results below:

\begin{itemize}

    \item When applied to a set of 2831 VFS galaxies with a k value of three, the algorithm divides our galaxies in this four-dimensional parameter space into three broad classes: dwarf galaxies (1785), spheroids (350), and large disks (696). While the three FCs overlap significantly on two-dimensional axes (Figure \ref{fig:faux_pca}), k-means clustering is able to robustly separate our galaxies into distinct groupings in four-dimensional feature space. Indeed, these FCs are structurally different and correlate with main sequence offset, \textit{NUV}-\textit{r} and W1-W3 colors, and general morphological T-type.

    \item The $\Delta$log(SFR) profiles of these FCs are such that the peaks of the dwarf and large disk galaxies are at and slightly above the main sequence line, with a tail for the dwarf galaxies extending to nearly -5 dex. The peak of the spheroid galaxies lies underneath the main sequence line by -2.5 dex. Each FC is a distinct $\Delta$log(SFR) population, and we can rule out with $>3\sigma$ significance that any of the three FCs are drawn from the same parent distribution.
    

    \item Evaluating the FC composition of galaxy environments from most to least dense, we find that spheroid galaxies in our sample our less common in low density environments and most common nearest to the Virgo cluster center. The broad environment representations for the dwarf and large disk galaxies, conversely, do not show clear trends.
    
    \item If we examine the $\Delta$log(SFR) distributions within each FC according to environment, we observe the general tendency of denser environments imposing stronger quenching effects on galaxies for a fixed FC. For the dwarf and large disk galaxies, SFRs are not strongly affected until the galaxies reach a rich group. For the spheroids, the SFR decreases smoothly moving to progressively dense environments. We verify that these trends do not coincide with trends of S\'ersic index for either W1 or \textit{g}-band, meaning that the morphology-density relation does not account for these differences in main sequence offset. We propose that these results indicate that the environment mechanisms instead act on a galaxy with varying efficiency depending on the galaxy's structural type. 


    \item We look at the role of satellite versus central labels from \citet{tempel2017}, both on their own and with respect to the local density characterizations using a fifth-nearest-neighbor treatment. We find that 2221 of our 2831 galaxy subset have RA-DEC matches with the \citet{tempel2017} catalog. The majority of our large disk galaxies are considered centrals while the spheroids and dwarf galaxies are primarily satellites. For all FCs, the satellite galaxies show a systematically lower SFR. Finely splitting the central and satellite populations for each FC into quantiles of local density, we find that any SFR differences between satellite and central galaxies disappear for the spheroid and large disks. The dwarf galaxies, however, continue to show more suppressed satellite galaxies in fourth quantile densities. There is a clear signal that the densest quantile suppresses SFR of satellite galaxies in all FCs compared to the less dense environments. 

    \item The median morphological T-types for each FC loosely map onto the expectation for each FC: the dwarf galaxies are primarily 7-8 (Sdm-Im), the spheroids closer to -2 (S0), and the extended disks around 5-6 (Sc-Scd) \citep{devaucouleurs1991book, makarov2014}. The T-types for dwarf galaxies show the most scatter. The dwarf galaxies with $T>$0 are more numerous and have noticeably different colors, star-forming main sequence offsets, and global environment. T$\le$0 galaxies are more common in dense environments, have redder \textit{NUV}-\textit{r} and are almost exclusively below the main sequence. 
    
    
\end{itemize}

This larger discussion of T-type subpopulations hidden in our FCs highlights that the four structural parameters we used in our k-means pipeline were not sufficient to resolve these subpopulations, instead grouping the Virgo galaxies into three broad classes. While this paper has focused on the utility of clustering algorithms to divide large samples of galaxies into these structurally similar classes, future work could consist of delving into whether additional measures of a galaxy's stellar structure could parse these subpopulations into their own FCs. As we caution in Section \ref{sec:ttype}, however, the selection of parameters which are directly linked to SFR or strongly manipulated by environmental processing conflicts with the goal of creating controlled galaxy populations. The testing of alternative parameters may also show that T-type does not perfectly encode a galaxy's underlying stellar structure -- and vice versa. 

The undercurrent objective of studying the applications of k-means clustering is the potential for upscaling to larger sweeps of the Universe. With the breadth of galaxy surveys delivering terabytes of data, there is an increasing need for automated galaxy characterizations that can accommodate many parameters. For our own work, we will be measuring the effects of global environment on star-forming disk sizes for a sample more than an order of magnitude larger than the VFS and encompassing more clusters than simply Virgo (WISESize; Conger et al., in prep.). The ability to generate groupings of galaxies with relatively fixed structural parameters, especially in a way that is objective and reproducible, is a valuable step to confront the high-dimensionality problem in galaxy evolution and better establish the relationships between a galaxy's ability to form stars and its place in the Cosmic Web.





\section{Acknowledgments}
The authors thank the International Space Sciences Institute (ISSI) in Bern, Switzerland who hosted collaboration meetings as part of the ISSI COSWEB team. They also thank the Max Kade Center at the University of Kansas for hosting a collaboration meeting organized by members of the Virgo Filaments Survey team. K.C. would like to acknowledge the support of the Maynard Redeker scholarship at the University of Kansas, as well as the Ralston Summer fellowship awarded by the University of Kansas Department of Physics \& Astronomy. G.R. and R.A.F. gratefully acknowledge support from NSF grants AST-1716657 and AST-2308127, and from a NASA ADAP grant 80NSSC21K0640.

The Legacy Surveys consist of three individual and complementary projects: the Dark Energy Camera Legacy Survey (DECaLS; Proposal ID \#2014B-0404; PIs: David Schlegel and Arjun Dey), the Beijing-Arizona Sky Survey (BASS; NOAO Prop. ID \#2015A-0801; PIs: Zhou Xu and Xiaohui Fan), and the Mayall z-band Legacy Survey (MzLS; Prop. ID \#2016A-0453; PI: Arjun Dey). DECaLS, BASS and MzLS together include data obtained, respectively, at the Blanco telescope, Cerro Tololo Inter-American Observatory, NSF’s NOIRLab; the Bok telescope, Steward Observatory, University of Arizona; and the Mayall telescope, Kitt Peak National Observatory, NOIRLab. Pipeline processing and analyses of the data were supported by NOIRLab and the Lawrence Berkeley National Laboratory (LBNL). The Legacy Surveys project is honored to be permitted to conduct astronomical research on Iolkam Du’ag (Kitt Peak), a mountain with particular significance to the Tohono O’odham Nation.

NOIRLab is operated by the Association of Universities for Research in Astronomy (AURA) under a cooperative agreement with the National Science Foundation. LBNL is managed by the Regents of the University of California under contract to the U.S. Department of Energy.

This project used data obtained with the Dark Energy Camera (DECam), which was constructed by the Dark Energy Survey (DES) collaboration. Funding for the DES Projects has been provided by the U.S. Department of Energy, the U.S. National Science Foundation, the Ministry of Science and Education of Spain, the Science and Technology Facilities Council of the United Kingdom, the Higher Education Funding Council for England, the National Center for Supercomputing Applications at the University of Illinois at Urbana-Champaign, the Kavli Institute of Cosmological Physics at the University of Chicago, Center for Cosmology and Astro-Particle Physics at the Ohio State University, the Mitchell Institute for Fundamental Physics and Astronomy at Texas A\&M University, Financiadora de Estudos e Projetos, Fundacao Carlos Chagas Filho de Amparo, Financiadora de Estudos e Projetos, Fundacao Carlos Chagas Filho de Amparo a Pesquisa do Estado do Rio de Janeiro, Conselho Nacional de Desenvolvimento Cientifico e Tecnologico and the Ministerio da Ciencia, Tecnologia e Inovacao, the Deutsche Forschungsgemeinschaft and the Collaborating Institutions in the Dark Energy Survey. The Collaborating Institutions are Argonne National Laboratory, the University of California at Santa Cruz, the University of Cambridge, Centro de Investigaciones Energeticas, Medioambientales y Tecnologicas-Madrid, the University of Chicago, University College London, the DES-Brazil Consortium, the University of Edinburgh, the Eidgenossische Technische Hochschule (ETH) Zurich, Fermi National Accelerator Laboratory, the University of Illinois at Urbana-Champaign, the Institut de Ciencies de l’Espai (IEEC/CSIC), the Institut de Fisica d’Altes Energies, Lawrence Berkeley National Laboratory, the Ludwig Maximilians Universitat Munchen and the associated Excellence Cluster Universe, the University of Michigan, NSF’s NOIRLab, the University of Nottingham, the Ohio State University, the University of Pennsylvania, the University of Portsmouth, SLAC National Accelerator Laboratory, Stanford University, the University of Sussex, and Texas A\&M University.

BASS is a key project of the Telescope Access Program (TAP), which has been funded by the National Astronomical Observatories of China, the Chinese Academy of Sciences (the Strategic Priority Research Program “The Emergence of Cosmological Structures” Grant \# XDB09000000), and the Special Fund for Astronomy from the Ministry of Finance. The BASS is also supported by the External Cooperation Program of Chinese Academy of Sciences (Grant \# 114A11KYSB20160057), and Chinese National Natural Science Foundation (Grant \# 12120101003, \# 11433005).
 
The Legacy Survey team makes use of data products from the Near-Earth Object Wide-field Infrared Survey Explorer (NEOWISE), which is a project of the Jet Propulsion Laboratory/California Institute of Technology. NEOWISE is funded by the National Aeronautics and Space Administration.

The Legacy Surveys imaging of the DESI footprint is supported by the Director, Office of Science, Office of High Energy Physics of the U.S. Department of Energy under Contract No. DE-AC02-05CH1123, by the National Energy Research Scientific Computing Center, a DOE Office of Science User Facility under the same contract; and by the U.S. National Science Foundation, Division of Astronomical Sciences under Contract No. AST-0950945 to NOIRLab.

\bibliography{report_template}

\appendix

\section{CIGALE Module Selection}
\label{appendix:cigale}

The setup of \texttt{CIGALE} begins with selecting modules which will govern the underlying physics of the SED models. We choose the following creation modules for the \texttt{CIGALE} run on our sample:
\begin{itemize}
    \item Treatment of star formation histories with a decaying exponential (\texttt{sfh2exp})
    \item \citet{bc03} library of single stellar populations (\texttt{bc03})
    \item Computation of of nebular emission (\texttt{nebular})
    \item \citet{cf00} modified dust attenuation law (\texttt{dustatt\_modified\_CF00})
    \item \cite{dl2014} templates for dust emission (\texttt{dl2014})
    \item \cite{skirtor2016} active galactic nuclei (AGN) templates (\texttt{skirtor2016})
    \item Redshifting of the model and computation of absorption due to the intergalactic medium (IGM; \texttt{redshifting})
\end{itemize}

The initialization file for \texttt{CIGALE} includes sections where users can either use the default range of initial guesses for each parameter comprising a module, or overwrite the default guesses with their own ranges. While generating model templates, \texttt{CIGALE} will create one SED model per parameter combination, meaning that computation time scales directly with the number of initial guesses. We populate these ranges in a way that balances the conflicting desires of a complete and physically-motivated sampling of parameter space, and reducing of total computation time to generate SED models.

\begin{table}
\centering
\caption{Tabulated modules and their initial parameters for our \texttt{CIGALE} SED fitting routine. Every SED fit to the photometric data of a galaxy uses some mix of these initial parameters until every combination is exhausted.}
\begin{tabular}{|c|c|c|}
\hline
Module & Parameter & Initial Guesses \\
\hline
sfh2exp  & tau\_main (Myr) & 300, 500, 1000, 3000, 6000, 10000 \\
  & tau\_burst (Myr) & 100, 200, 400 \\
  & f\_burst & 0, 0.001, 0.005, 0.01, 0.05, 0.1 \\
  & age (Myr) & 1000, 3000, 5000, 7000, 10000, 13000 \\
  & burst\_age (Myr) & 20, 80, 200, 400, 800, 1000 \\
  & sfr\_0 (Msun/yr) & 1.0 \\
  & normalize & True \\
\hline
bc03 & imf & 1 (Chabrier) \\
  & metallicity & 0.004, 0.02, 0.05 \\
  & separation\_age (Myr) & 10 \\
\hline
nebular & logU & -2.0 \\
  & zgas & 0.02 \\
  & ne & 100 \\
  & f\_esc & 0.0 \\
  & f\_dust & 0.0 \\
  & lines\_width (km/s) & 300 \\
  & emission & True \\
\hline
dustatt\_modified\_CF00 & Av\_ISM & 0.0, 0.01, 0.025, 0.03, 0.035, 0.04, \\
  & & 0.05, 0.06, 0.12, 0.15, 1.0, 1.3, \\
  & & 1.5, 1.8, 2.1, 2.4, 2.7, 3.0, 3.3 \\
  & mu & 0.44 \\
  & slope\_ISM & -0.7 \\
  & slope\_BC & -1.3 \\
  & filters & V\_B90 \& FUV \\
\hline
dl2014 & qpah & 2.5 \\
  & umin & 1.0, 5.0, 10.0 \\
  & alpha & 1.0, 2.0, 2.8 \\
  & gamma & 0.02, 0.1 \\
\hline
skirtor2016 & t & 7.0 \\
  & pl & 1.0 \\
  & q & 1.0 \\
  & oa (deg) & 40.0 \\
  & R & 20.0 \\
  & Mcl & 0.97 \\
  & i (deg) & 30.0 \\
  & disk\_type & 1 (Schartmann (2005)) \\
  & delta & -0.36 \\
  & fracAGN & 0.0, 0.05, 0.1, 0.5 \\
  & lambda\_fracAGN & 0/0 \\
  & law & 0 (SMC) \\
  & EBV (mag) & 0.03 \\
  & temperature (K) & 100 \\
  & emissivity & 1.6 \\
\hline
\end{tabular}
\label{tab:cigale_parameters}
\end{table}

\newpage

\section{K-Means Stability Tests}
\label{appendix:stability}

One approach to optimizing k is to apply the silhouette method. Formally, the silhouette score is defined as 

\begin{center}
    \begingroup
        \Large
            \begin{math}
                 s = \frac{b-a}{max(b,a)},
            \end{math}
    \endgroup
\end{center}
where \textit{a} is the mean intra-cluster distance (the distance between a cluster's centroid and its constituent data points) and \textit{b} is the distance between a cluster and its nearest neighbor cluster. The score thus ranges from -1.0 to 1.0, where $<0$ indicates poor clustering, 0 indicates overlapping clusters, and values closer to 1 correspond to a more robustly defined set of FCs.

\begin{figure}[h]
\includegraphics[scale=0.5]{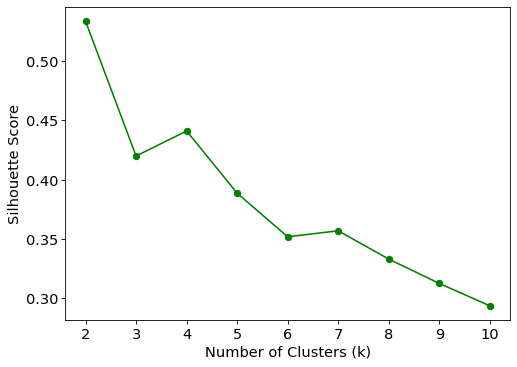}
\centering
\caption{A visualization of the silhouette score versus the number of k-means clusters for our sample set. A higher score signals a more well-separated clustering result.}
\vspace{.5mm}
\label{fig:silhouette}
\end{figure}

In Figure \ref{fig:silhouette}, we plot the resultant silhouette scores for a variety of test k values. While none of the scores dip below 0.30, there is almost a monotonic decline once the number of clusters moves beyond 4, suggesting that partitioning of the data with k$\ge$5 creates FCs that are less distinct. The plot also suggests that the division of our galaxy sample into 2 or 4 FCs yields the most stable configuration. The k=2 case is highest, signaling that two broad morphology classes adequately describe our galaxy population; however, a silhouette score difference of $<0.1$ is not a sufficient criterion to argue in preference for one k value over another. Moreover, as we have shown in this paper, a simple k=2 split fails to capture the more meaningful class(es) that emerges when instead partitioning into 3 FCs.


We use a second quantitative measure, the Adjusted Rand Index (ARI), as a complementary measure of robustness for our test k values. ARI is a statistical metric of the consistency of cluster assignments across different initializations, corrected for chance, where a single run corresponds to a different random seed. To compare two runs, the algorithm will count how many galaxies had the same clustering assignment between runs and how many differed. The index, also ranging from -1.0 to +1.0, indicates whether every run is indistinguishable (+1.0), comparable to random chance assignments (0), or that the comparisons are significantly \textit{worse} than random (-1.0). A high ARI therefore translates to a more stable and reproducible k-means partitioning.

We calculate the ARI scores for a set of 100 runs per k value, and tabulate the mean and standard deviation for these sets in Table \ref{tab:ari}. In addition to a higher ARI score, a small standard deviation is indicative of a robust k value, since there is minimal variability in cluster assignments. From this Table, we see tjat the k=3 case (mean=0.993, $\sigma=0.006$) shows the highest stability and lowest variance, suggesting that it is the most insensitive of our test cases to initialization parameters. While k=2 is nearly identical in its ARI statistics, the k=2 case partitions galaxies according to physical size and thus does not offer any new information. This finding reinforces our choice of k=3 as the optimal k-means partitioning of our galaxy population.

\begin{table}
\centering
\caption{Adjusted Rand Index (ARI) statistics as a function of the number of clusters $k$. The mean and standard deviation are computed after 100 runs with different random seeds.}
\begin{tabular}{|ccc|}
\hline
$k$ & Mean ARI & Standard Deviation ($\sigma$) \\
\hline
2  & 0.991 & 0.009 \\
3  & \textbf{0.993} & \textbf{0.006} \\
4  & 0.744 & 0.226 \\
5  & 0.897 & 0.190 \\
6  & 0.694 & 0.216 \\
7  & 0.719 & 0.176 \\
10 & 0.656 & 0.116 \\
\hline
\end{tabular}
\label{tab:ari}
\end{table}




\end{document}